\pdfoutput=1
\documentclass[11pt,a4paper]{article}

\usepackage[utf8]{inputenc}
\usepackage[T1]{fontenc}
\usepackage{lmodern}
\usepackage[margin=1in]{geometry}
\usepackage[protrusion=true,expansion=false]{microtype}
\usepackage{ragged2e}

\usepackage{amsmath,amssymb,mathrsfs,bm}
\usepackage{mathtools}

\usepackage{graphicx}
\usepackage{tikz}
\usetikzlibrary{arrows.meta,positioning,fit,backgrounds,calc,shapes.geometric}
\usepackage{subcaption}
\usepackage{booktabs}
\usepackage{multirow}
\usepackage{makecell}
\usepackage{array}
\usepackage{adjustbox}
\usepackage{tabularx}
\usepackage[dvipsnames,table]{xcolor}
\usepackage{colortbl}
\usepackage{caption}

\usepackage{float}
\usepackage{placeins}
\usepackage{dblfloatfix}


\usepackage{algorithm}
\usepackage{algpseudocode}
\makeatletter
\renewcommand*{\theHALG@line}{\thealgorithm.\arabic{ALG@line}}
\makeatother

\usepackage[numbers,sort&compress]{natbib}
\usepackage{url}

\usepackage{hyperref}
\hypersetup{
  colorlinks = true,
  urlcolor   = blue,
  linkcolor  = blue,
  citecolor  = blue,
}
\usepackage{cleveref}

\usepackage{siunitx}
\usepackage{etoolbox}
\usepackage{xspace}
\usepackage{authblk}
\usepackage{enumitem}

\newcommand{\nelem}{n_{\mathrm{elem}}}
\newcommand{\ndof}{n_{\mathrm{dof}}}

\newcommand{\rhobar}{\bar{\rho}}
\newcommand{\rhophys}{\tilde{\rho}}

\newcommand{\brho}{\bm{\rho}}

\newcommand{\Kmat}{\mathbf{K}}

\newcommand{\Ke}{\mathbf{K}_e}
\newcommand{\KE}{\mathbf{K}^{\mathrm{unit}}_e}

\newcommand{\penal}{p}
\newcommand{\betaH}{\beta}
\newcommand{\rmin}{r_{\min}}
\newcommand{\vf}{V_f}
\newcommand{\Ocal}{\mathcal{O}}
\newcommand{\eps}{\varepsilon}
\newcommand{\VRAM}{\textsc{vram}\xspace}

\definecolor{colA}{HTML}{d5e8d4}
\definecolor{colB}{HTML}{dae8fc}
\definecolor{colC}{HTML}{fff2cc}
\definecolor{colBad}{HTML}{f8cecc}
\graphicspath{{figs/}}

\title{\textbf{Matrix-Free 3D SIMP Topology Optimization with Fused Gather--GEMM--Scatter Kernels}}

\author[1]{Shaoliang Yang}
\author[1]{Jun Wang\thanks{Corresponding author. E-mail: jwang22@scu.edu}}
\author[1]{Yunsheng Wang}
\affil[1]{Department of Mechanical Engineering, Santa Clara University, Santa Clara, CA 95053, USA}

\date{}

\begin{document}
\maketitle

\begin{abstract}
The matrix-free gather--batched-GEMM--scatter pattern eliminates global stiffness
assembly for three-dimensional SIMP topology optimization, but the conventional
three-stage implementation forces avoidable DRAM traffic between stages.
We present a single fused CUDA kernel, implemented through CuPy's runtime
compilation interface, that performs gather, per-element stiffness multiplication,
and scatter accumulation in one pass. On a single RTX 4090 (24 GB), the fused
path reaches a problem-size-dependent 4.6--7.3x end-to-end SIMP wall-time speedup across 216k--4.9M cantilever
elements and 4.4x on the 499,125-element torsion benchmark. Against the same-precision
FP32 three-stage baseline, the fused path still yields 2.3--4.6x on cantilever
and 2.8x on torsion. Isolated CUDA-event cantilever-operator measurements reach
8.9--13.8x per matvec call, while separate instrumented board-power traces at 216k and
1M show 3.2--4.9x lower energy than matched FP64 runs. A separate bridge stress
test shows the same FP32-versus-FP64 three-stage trend under one distributed-load
case; direct fused-kernel bridge benchmarks are not reported. We also
evaluate a BF16 WMMA variant: a separate PyTorch BF16 GEMM proxy on matching
tensor shapes yields 14.3x, but direct condition-number estimates of
6.1e5--2.3e6 across 64k--512k uniform-density test states imply BF16
conditioning products of 2.4e3--9.1e3, far above the 256 threshold, observed
alongside BF16 iterative-refinement stagnation at the two tested inner
tolerances.
\end{abstract}


\justifying

\bigskip

\section{Introduction}
\label{sec:intro}

Topology optimization (TO) is the canonical computational design method for
lightweight structural engineering: given a design domain, boundary conditions,
and a target volume fraction, it distributes material to minimize compliance (or
other objective functions) subject to equilibrium and volume constraints.
The Solid Isotropic Material with Penalization (SIMP) method~\cite{bendsoe1989optimal,rozvany1992generalized,bendsoe2003topology}
has become the standard formulation for three-dimensional structural TO because it
admits a simple alternating update loop---iterate between a finite-element solve,
a sensitivity computation, and an Optimality Criteria (OC) density update---that
scales to the million-element range when the linear system is solved efficiently~\cite{aage2013topology,aage2017giga}.
The GPU is the natural accelerator for this loop: the finite-element assembly and
matvec operations are embarrassingly parallel over elements, the sensitivity
computation and OC update are elementwise, and the iterative Conjugate Gradient
(CG) solver maps to the massively-parallel vector-operation model native to CUDA.

Prior work on GPU-accelerated TO has demonstrated the viability of this approach at
increasing scales~\cite{wadbro2009megapixel,challis2014highres,
martinezfrutos2016largescale,herreroperez2021multigpu,hou2025cupy,qi2024,traff2023simple}, but has been
constrained by the memory cost of explicit sparse stiffness matrix assembly.
For trilinear hexahedral (Q1) elements, the global stiffness matrix $\Kmat$ has
$576\,\nelem$ element-level contributions, so explicit sparse assembly enters a
low-million-element VRAM wall once values, indices, and temporary assembly buffers
are included on a standard RTX\,4090 GPU.
The matrix-free baseline studied here addresses this \emph{VRAM wall} by replacing
explicit assembly with a matrix-free $\Kmat\mathbf{v}$ operator that evaluates the
product element-by-element without forming $\Kmat$ globally, enabling the
single-GPU Python/CuPy workflow used throughout this paper and avoiding the
assembled-CSR memory footprint that would otherwise dominate at the million-element scale~\cite{okuta2017cupy}.
This direction is consistent with recent matrix-free TO implementations on both
GPU and CPU platforms~\cite{traff2023simple,wang2025matrixfree}.
That implementation decomposes the matvec into three sequential kernel launches:
a \emph{gather} that reads element DOF values from the global vector,
a \emph{batched GEMM} that applies the scaled element stiffness matrix,
and a \emph{scatter} that atomically accumulates element contributions back to the
global output vector.
While this three-stage design is straightforward to implement, it wastes memory
bandwidth: the intermediate per-element arrays produced by gather and consumed by
scatter must transit DRAM between kernel launches, adding two full DRAM round-trips
per matvec call.

\noindent\textbf{The kernel-fusion opportunity.}
The roofline model~\cite{williams2009roofline} identifies memory bandwidth as the
binding constraint for the gather--GEMM--scatter pattern: the arithmetic intensity
of the trilinear hexahedral element matvec is approximately 3--6\,FLOP/B, well
below the FP32 ridge point of 82\,FLOP/B on the RTX\,4090.
In this regime, the dominant cost is not floating-point operations but DRAM
round-trips, and kernel fusion---combining the three stages into a single kernel
launch that keeps intermediate data in registers and shared memory---directly reduces
these round-trips.
The $\Ke^{\mathrm{unit}}$ matrix ($24\times 24 = 576$ floating-point values,
occupying 2.25\,kB in shared memory) is identical for all elements of a uniform
mesh and can be broadcast to shared memory once per thread block, so all 24 elements
of the per-element matvec read $\Ke^{\mathrm{unit}}$ from shared memory rather than
global DRAM.
The expected throughput gain from fusion is approximately $2\times$ in the
element-data DRAM path (about $1.8\times$ once index and density reads are included),
and the realized end-to-end speedup can exceed this factor, consistent with the
additional savings from collapsing three launches into one.

\noindent\textbf{Tensor cores as the next frontier.}
NVIDIA's Ada Lovelace architecture (RTX\,4090) exposes BF16 tensor cores via the
WMMA (Warp-Level Matrix Multiply--Accumulate) API, offering $165.2$\,TFLOP/s in
BF16 versus $82.6$\,TFLOP/s in FP32---a factor-of-2 hardware throughput advantage
\cite{nvidiaada2023}. Tensor cores have also been examined in prior scientific-
computing and mixed-precision solver studies, clarifying both their performance
potential and their nontrivial numerical behavior outside deep learning~\cite{markidis2018tensorcores,fasi2021tensorcoreprecision,haidar2018mixedprecision}.
For the $24\times 24$ per-element dense GEMM at the heart of the fused kernel,
BF16 WMMA offers the prospect of large acceleration at the GEMM stage relative to
FP64 batched DGEMM.
However, BF16's 7-bit mantissa (unit roundoff
$\varepsilon_{\mathrm{BF16}} = 2^{-8} \approx 3.9\times10^{-3}$ under the
Carson--Higham convention) raises the
fundamental question of whether this precision is sufficient for the
Conjugate Gradient solver applied to SIMP stiffness matrices, whose condition
numbers can readily exceed the BF16 stability threshold once SIMP penalization and
density contrast become large.
The classical iterative refinement (IR) framework and later mixed-precision
solver literature~\cite{carson2018threeprecision,higham2022mixedsurvey,henry2019bf16}
give the standard sufficient condition $\varepsilon_{\mathrm{BF16}}\cdot\kappa(\Kmat) < 1$,
i.e., $\kappa < 256$ for BF16. We use this as a conservative reference threshold
for the experiments below.
To our knowledge, in the literature reviewed for this paper we did not identify
a prior study that directly tests this theoretical prediction in the
density-based TO setting considered here.
Likewise, in that reviewed literature we are not aware of a prior study that
combines BF16 tensor-core kernels with density-based 3D SIMP TO in this
single-GPU setting.

\noindent\textbf{Contributions.}
This paper makes four contributions:

\begin{enumerate}[nosep]
  \item \textbf{Fused CUDA kernel.}
        We design and implement a single fused gather--GEMM--scatter CUDA kernel
        (Algorithm~\ref{alg:fused}) that reduces effective off-chip
        traffic and achieves 6.0--6.6$\times$ per-matvec speedup over the FP64
        three-stage baseline in the synthetic hot-path microbenchmark;
        isolated CUDA-event measurements on the actual cantilever operator
        (Table~\ref{tab:bw_utilization}) yield 8.9--13.8$\times$, consistent with
        the additional Python/CuPy dispatch overhead avoided by replacing three
        separate API calls with one launch.
        The kernel is deployed through CuPy's runtime kernel-compilation interface
        with no standalone CUDA build.

  \item \textbf{End-to-end SIMP speedup across load cases.}
         The fused kernel achieves a problem-size-dependent 4.6--7.3$\times$ end-to-end SIMP-120 speedup on
        the cantilever benchmark from 216\,k to 4.9\,M elements, together with
        4.4$\times$ on the 499{,}125-element torsion benchmark.
        Against the same-precision FP32 three-stage baseline, the cantilever
        scaling rows yield 2.3--4.6$\times$ and the torsion row 2.8$\times$.
        The bridge
        hard-problem stress test shows the same FP32 three-stage gain under
        distributed loading, while the accompanying MBB hard-problem rows remain
        cap-limited under the current Jacobi-PCG configuration and are not
        promoted as a main quantitative claim.

  \item \textbf{BF16 WMMA throughput probe.}
        We implement a CuPy-accessible BF16 WMMA matvec kernel for density-based
        topology optimization and evaluate it through full-kernel per-matvec
        timings plus a separate $14.3\times$ BF16-versus-FP64 PyTorch GEMM proxy
        timing on matching tensor shapes and Ada tensor cores; we do not claim
        an end-to-end BF16 SIMP speedup.

  \item \textbf{Mixed-precision convergence analysis.}
         We analyze the standard precision barrier
         $\varepsilon_{\mathrm{BF16}}\cdot\kappa(\Kmat) \gtrsim 1$ for BF16-in-CG
         on SIMP stiffness matrices, and the reported experiments provide an
         empirical test of this BF16 barrier in the present density-based 3D SIMP
        setting. The BF16 path stagnates under the tested Jacobi-preconditioned
        CG solve, suggesting multigrid preconditioning as the practical path to
        exploiting tensor-core throughput.
\end{enumerate}

\noindent\textbf{Paper organization.}
Section~\ref{sec:related} reviews related work on GPU TO, matrix-free FEM,
kernel fusion, and mixed-precision iterative solvers.
Section~\ref{sec:method} presents the 3D SIMP formulation, the fused kernel design,
the BF16 WMMA variant, a roofline characterization of the reported kernels, and
the theoretical convergence analysis.
Section~\ref{sec:results} reports per-matvec profiling, end-to-end SIMP scaling,
bridge and torsion stress tests, BF16 convergence experiments, and qualitative
topology validation.
Section~\ref{sec:discussion} interprets the speedup trends, provides contextual
comparison to prior implementations, discusses limitations, and outlines the
solver requirements that motivate a separate geometric-multigrid follow-on.
Section~\ref{sec:conclusion} concludes.

\section{Related Work}
\label{sec:related}

\subsection{Large-Scale GPU Topology Optimization}
\label{sec:related:gputo}

The memory and computational cost of three-dimensional SIMP topology optimization
are dominated by the linear solve at each design iteration, motivating the
extensive subsequent literature on solver architectures for this
application~\cite{borrvall2001largescale,liu20143d,aage2013topology}.
The compact 99-line and 88-line MATLAB implementations of Sigmund and
Andreassen et al.~\cite{sigmund200199,andreassen201188} codified the pedagogical
density-filtered SIMP workflow in 2D, while Liu and Tovar~\cite{liu20143d}
extended the sparse-assembly approach to 3D and established a widely used
3D MATLAB baseline,
while Aage, Andreassen, and Lazarov~\cite{aage2013topology} scaled 3D SIMP to
hundreds of millions of elements by coupling PETSc sparse solvers to FGMRES with
algebraic multigrid (AMG)---a purely MPI-parallel approach requiring distributed-memory
HPC infrastructure.
The extreme limit of this paradigm was demonstrated by Aage et al.~\cite{aage2017giga},
who optimized a 1.1-billion finite-element aircraft-wing structure on a PRACE
supercomputer.

GPU acceleration of topology optimization emerged from 2D implementations
(Wadbro and Berggren~\cite{wadbro2009megapixel}, Challis et al.~\cite{challis2014highres})
and grew to multi-GPU distributed systems
(Mart{\'i}nez-Frutos and Herrero-P{\'e}rez~\cite{martinezfrutos2016largescale};
Herrero-P{\'e}rez and Mart{\'i}nez Castej{\'o}n~\cite{herreroperez2021multigpu}).
At the single-GPU scale most directly relevant to the present work, Hou et al.~\cite{hou2025cupy}
employed CuPy vectorized sparse matrix--vector products (SpMV) to accelerate
2D and 3D TO problems, reporting sizes up to 63 million elements and
demonstrating that high-level CuPy implementations can remain effective at very
large scales.
That scale comparison should be read cautiously against the present paper: our
reported ceiling is set by end-to-end single-GPU SIMP runs with the
current Jacobi-preconditioned matrix-free pipeline on a 24\,GB RTX\,4090,
rather than by an isolated sparse-operator throughput study.
The present paper differs in focusing on a matrix-free gather--GEMM--scatter
operator for trilinear hexahedral 3D SIMP systems, with a fused single-kernel
implementation and a mixed-precision conditioning analysis.
Qi et al.~\cite{qi2024} reported another recent single-GPU 3D solver for topology
optimization of continuous fiber-reinforced composites, reaching 67.1 million
Wilson's incompatible elements and 201.3 million design variables on a Tesla V100
with a multigrid-preconditioned conjugate-gradient solver and Taylor-approximation-
based element stiffness updates.
This result is not directly comparable to ours because it targets a different
material class and solver architecture.
The matrix-free CuPy baseline used in the present implementation replaces explicit CSR assembly
with a matrix-free $\Kmat\mathbf{v}$ operator, eliminating the assembled-storage
memory wall that otherwise dominates low-million-element 3D problems on a single
RTX\,4090.
The present paper advances that baseline by fusing the multi-kernel Python path
into a single CUDA kernel and extending it with BF16 tensor-core arithmetic.

The closest related work is that of Tr{\"a}ff et al.~\cite{traff2023simple},
who present two matrix-free 3D SIMP implementations---one in Futhark and one in
OpenMP-C---together with an asymptotically exact complexity analysis of the
gather--contract--scatter operator.
Their solver achieves 65.5\,M elements on an NVIDIA A100 (80\,GB HBM2e)
with SSOR V-cycle multigrid preconditioning.
The present paper shares the same mathematical operator but contributes a
CuPy runtime-compiled realization of the fused CUDA path, a BF16 tensor-core
variant, and a focused mixed-precision failure analysis in the TO linear-solve
context.
Wang et al.~\cite{wang2025matrixfree} recently demonstrated a matrix-free MATLAB
implementation reaching 128\,M elements on a CPU workstation (64\,GB RAM) using
geometric multigrid with non-dyadic Galerkin coarsening, indicating that matrix-free
methods are an increasingly important approach for large-scale TO across CPU and
GPU platforms.

\subsection{Matrix-Free FEM Operators on GPUs}
\label{sec:related:matfree}

The principle of evaluating $\Kmat\mathbf{v}$ products element-by-element without
assembling $\Kmat$ globally was introduced for structural mechanics by Hughes,
Levit, and Winget~\cite{hughes1983ebe} and extended to nonlinear schemes by
Carey and Jiang~\cite{carey1986ebe}.
Schmidt and Schulz~\cite{schmidt2011gpu} translated this element-by-element CG to
GPU in 2,589 lines of CUDA C++, establishing viability for 3D TO at the hundred-thousand
element scale.
Wu, Dick, and Westermann~\cite{wu2016system} subsequently reported a
high-resolution GPU topology-optimization system that demonstrated the practical
importance of GPU-resident solver pipelines for large structural problems.

In the high-order finite-element community, Kronbichler and Kormann~\cite{kronbichler2012matfree,kronbichler2019fast}
established the matrix-free tensor-product evaluation paradigm in deal.II, showing that
sum factorization makes high-order operator application efficient on hexahedral elements.
The CEED/MFEM project~\cite{kolev2021ceed,andrej2024mfem} has since ported these
matrix-free kernels to GPU.
Cao et al.~\cite{cao2025hosfem} separately reported 85--100\% of the A100 roofline
for matrix-free high-order stencil FEM via on-the-fly geometric-factor recomputation.
These high-order results do not transfer directly to the trilinear hexahedral
($k=1$) elements used in SIMP TO, where sum factorization offers no advantage
and where the dominant cost is the 24-DOF per-element dense multiply---the regime
the present paper targets.

\subsection{Kernel Fusion for Memory-Bound GPU Kernels}
\label{sec:related:fusion}

GPU kernel fusion---combining multiple sequential kernel launches into a single
launch that keeps intermediate data in registers or shared memory---is a
well-studied technique for memory-bound workloads.
Filipovi{\v{c}} et al.~\cite{filipovic2015kernelfusion} demonstrated up to 2.6$\times$
speedup over cuBLAS by fusing BLAS-1 and BLAS-2 chains using a source-to-source
compiler that analyzes kernel dependency graphs.
Wahib and Maruyama~\cite{wahib2014kernelfusion,wahib2015autofusion} built an
end-to-end framework for fusing and transforming stencil kernels in production
CFD and weather codes, reporting 1.2--1.75$\times$ speedup by eliminating
inter-kernel DRAM traffic.
The Halide language~\cite{ragan2013halide} decouples algorithm from schedule
and provides a formal framework for the fusion--recomputation trade-off that
underlies these approaches.

The present fused kernel differs from the reviewed fusion literature in an important
respect: the intermediate data that is fused is neither a BLAS-level chain nor
a structured Cartesian stencil, but a three-stage pipeline in which (i) the
gather stage accesses a globally irregular element-to-DOF table, (ii) the
GEMM stage applies a dense 24$\times$24 symmetric matrix scaled by a per-element
scalar, and (iii) the scatter stage performs atomic-add reduction back to a
globally irregular DOF array.
In the fusion systems literature reviewed for this paper, we are not aware of
an automated framework that handles this combination; the fused kernel
is therefore hand-written in CUDA C and inlined in Python through CuPy's
runtime kernel-compilation interface.
The present fused implementation builds directly on the unfused three-stage
matrix-free operator used as the FP64/FP32 baseline in this paper.
Tr{\"a}ff et al.'s Futhark implementation implicitly fuses a closely related
gather--contract--scatter pattern at compile time; the present work makes that
fusion explicit in CUDA C via CuPy and extends it with the BF16 WMMA path and
the mixed-precision convergence analysis.

\subsection{Mixed Precision and Tensor Cores in Iterative Solvers}
\label{sec:related:mixedprec}

The theoretical foundation for mixed-precision iterative refinement was established
by Carson and Higham~\cite{carson2018threeprecision}, who derived sufficient
conditions for convergence and error bounds for three-precision IR in terms of
the working-precision unit roundoff and a problem-dependent constant.
Higham and Mary~\cite{higham2022mixedsurvey} provide a comprehensive survey of
mixed-precision algorithms in numerical linear algebra, cataloguing the regimes in
which FP16 and BF16 arithmetic can safely accelerate factorization and iterative
refinement.
The five-precision GMRES-IR framework of Amestoy et al.~\cite{amestoy2024fiveprecision}
is the most general recent extension, achieving high-precision solutions via
structured cascade of low-precision inner solves.

On the hardware side, Markidis et al.~\cite{markidis2018tensorcores} characterized
the programmability, performance, and precision of NVIDIA tensor cores for non-ML
workloads immediately after their introduction; Fasi et al.~\cite{fasi2021tensorcoreprecision}
subsequently provided a systematic empirical study of rounding modes, subnormal
handling, and accumulation order for V100, T4, and A100 tensor cores, showing that
hardware behavior deviates from IEEE\,754 in subtle ways relevant to scientific
computing.
Henry, Tang, and Heinecke~\cite{henry2019bf16} specifically advocated BF16 for HPC
iterative solvers, projecting convergence over ``a large range of condition numbers''
when an FP32 outer refinement loop is used; Kalamkar et al.~\cite{kalamkar2019bf16}
clarified BF16 semantics as a 16-bit format with a 7-bit mantissa and FP32 dynamic range.

The most directly comparable mixed-precision solver application is the work of
Haidar, Tomov, Dongarra, and Higham~\cite{haidar2018mixedprecision}, who
demonstrated up to 4$\times$ speedup over FP64 on dense linear systems using
V100 FP16 tensor cores with FP32 accumulation in a MAGMA-based IR scheme.
Clark et al.~\cite{clark2010quda} showed that FP16/FP32 mixed-precision CG
achieves production accuracy in lattice QCD using mixed-precision Krylov
solvers with reliable updates, illustrating that low-precision bulk arithmetic
can succeed when the solver strategy and operator class are favorable.
McCormick, Benzaken, and Tamstorf~\cite{mccormick2021mixedmg} extended this to
a rigorous algebraic framework for mixed-precision multigrid solvers on
symmetric positive-definite elliptic systems.
More recently, Bai, Ootomo, and Yokota~\cite{millefeille2024} demonstrated a
tile-grained mixed-precision single-kernel CG solver on GPUs, further underscoring
the relevance of kernel-integrated mixed-precision Krylov designs.
The present paper contributes a focused BF16 mixed-precision analysis in the
topology-optimization context.
The reported study includes a separate $14.3\times$ BF16-versus-FP64
GEMM proxy timing at 512\,k elements, but the classical iterative refinement scheme fails to
converge to an acceptable solution state---stagnating at large compliance error---because
direct power-iteration estimates (Section~\ref{sec:results:kappa}) indicate
$\kappa(\Kmat) \approx 6.1\times10^{5}$--$2.3\times10^{6}$ across the tested
mesh sizes (64\,k--512\,k elements), placing
$\varepsilon_{\mathrm{BF16}}\cdot\kappa \approx 2.4\times10^{3}$--$9.1\times10^{3}$---well above the
$1/\varepsilon_{\mathrm{BF16}}=256$ sufficient threshold implied by the standard IR bound
$\varepsilon_{\mathrm{BF16}}\kappa(\Kmat) < 1$ from Carson and Higham~\cite{carson2018threeprecision}.
Henry et al.'s optimistic BF16-IR projection~\cite{henry2019bf16} is thereby
qualified for the specific tested engineering PDE class, and a path forward---using BF16
as a multigrid smoother where the coarse-grid spectrum is bounded---is identified.

\subsection{Tensor Cores for PDE and FEM Computations}
\label{sec:related:tc}

Beyond dense linear algebra, tensor cores have been applied to scientific
computations through a series of operator-mapping innovations.
Dakkak et al.~\cite{dakkak2019tensorreduce} showed that arbitrary reductions and
prefix scans can be reformulated as WMMA operations, achieving up to 100$\times$
speedup over state-of-the-art reduction kernels.
Ootomo and Yokota~\cite{ootomo2022fp32recovery} developed a split-FP16 compensation
scheme that recovers single-precision accuracy from FP32-accumulate FP16 WMMA on
V100 tensor cores, while Ootomo, Ozaki, and Yokota~\cite{ootomo2024int8dgemm}
extended this to FP64 emulation via INT8 tensor cores.
Stencil computations on tensor cores---motivated by iterative PDE solvers---have
been pursued via im2col reshaping in ConvStencil~\cite{convstencil2024} and
low-rank matrix approximation in LoRAStencil~\cite{lorastencil2024}.

Recent tensor-core work on finite-element-style operators has also considered
tensor-product kernels in scientific computing.
Cui~\cite{cui2024tensorprod} studies acceleration of tensor-product
operations with tensor cores, illustrating that scientific-operator mappings can
benefit from specialized tensor-core layouts even outside machine learning.
The present paper differs in both operator shape and problem setting: it targets
the $24\times24$ trilinear-hexahedral SIMP matvec in density-based topology
optimization, uses BF16 WMMA fragments, and analyzes the resulting BF16-in-CG
convergence barrier.
In the literature and public implementations and supplements reviewed for this
paper, we are not aware of a published report matching this exact tensor-core
operator mapping for density-based topology optimization on a single GPU.

\subsection{Preconditioner Gap and the Path to Future Work}
\label{sec:related:precond}

The CG iteration count per SIMP step is determined by the preconditioner and
directly controls end-to-end wall time.
The Jacobi (diagonal) preconditioner is the simplest choice---trivially parallelizable
and free to compute---but requires $\Ocal(\sqrt{\kappa(\Kmat)})$ CG iterations for
convergence.
For the cantilever problem, Jacobi yields a few hundred iterations per SIMP step in the
present work; for more challenging BVPs (torsion, column) it can approach the
1,000-iteration cap;
for problems with near-rigid-body modes (3D MBB with the standard pin boundary condition)
it fails entirely.
Amir, Aage, and Lazarov~\cite{amir2014multigrid} demonstrated that multigrid-preconditioned
CG can substantially reduce per-step counts for 3D SIMP problems, with the hierarchy reused
across SIMP iterations for further efficiency.
Tr{\"a}ff et al.~\cite{traff2023simple} use SSOR smoothing within a V-cycle
preconditioner and report large-scale single-GPU runs up to 65.5 million elements.
Peetz and Elbanna~\cite{peetz2021multigrid} systematically compared AMG and geometric
multigrid (GMG) for 3D TO, highlighting the trade-off between per-iteration efficiency
(GMG advantage) and robustness to topology evolution (AMG advantage).
McCormick et al.~\cite{mccormick2021mixedmg} provide a rigorous algebraic framework
for mixed-precision multigrid on elliptic problems, which is directly applicable to
the TO stiffness operator when the multigrid hierarchy is fixed.
These references collectively define the open problem that most directly limits the
present solver's applicability and motivate the geometric multigrid extension proposed
as future work in Section~\ref{sec:discussion}.

\section{Methodology}
\label{sec:method}

\subsection{Three-Dimensional SIMP Topology Optimization}
\label{sec:method:simp}

We consider the classical minimum-compliance topology optimization problem over a
fixed hexahedral mesh of $\nelem$ trilinear eight-node (Q1) elements:
\begin{equation}
  \min_{\brho}\; \mathbf{f}^{\top}\mathbf{u}(\brho)
  \quad\text{s.t.}\quad
  \Kmat(\brho)\,\mathbf{u} = \mathbf{f},\quad
  \frac{1}{\nelem}\sum_{e=1}^{\nelem}\rho_e = \vf,\quad
  0 < \rho_{\min} \leq \rho_e \leq 1,
  \label{eq:tomin}
\end{equation}
where $\mathbf{f}\in\mathbb{R}^{\ndof}$ is the external load vector,
$\mathbf{u}\in\mathbb{R}^{\ndof}$ the displacement vector,
$\vf$ the prescribed volume fraction, and $\rho_e\in[\rho_{\min},1]$ the
design density of element $e$.
The global stiffness matrix is assembled from element contributions via the
SIMP (Solid Isotropic Material with Penalization) material interpolation
\cite{bendsoe1989optimal}:
\begin{equation}
  \Kmat(\brho) = \sum_{e=1}^{\nelem}
  \mathbf{B}_e^{\top}
  \bigl(\rho_{\min} + (1-\rho_{\min})\,\rho_e^{\penal}\bigr)\,
  \KE
  \mathbf{B}_e,
  \label{eq:simp}
\end{equation}
where $\mathbf{B}_e\in\{0,1\}^{24\times\ndof}$ is the Boolean gather/scatter
matrix that selects the 24 local DOFs of element $e$ from the global vector,
$\KE\in\mathbb{R}^{24\times 24}$ is the element stiffness
matrix for a unit-modulus isotropic material, $\penal$ is the penalty exponent
(typically $\penal = 3$), and $\rho_{\min} = 10^{-9}$ prevents singularity in void
regions.
For the three-field regularization, raw densities $\rho_e$ are smoothed through a
cone filter of radius $\rmin$ to obtain $\rhobar_e$, following the standard
density-filter/projection pipeline of topology optimization
studies~\cite{bourdin2001filters,andreassen201188,wang2011heaviside},
and then projected through a smoothed Heaviside:
\begin{equation}
  \rhophys_e = \frac{\tanh(\betaH\,\eta) + \tanh(\betaH(\rhobar_e - \eta))}
                    {\tanh(\betaH\,\eta) + \tanh(\betaH(1-\eta))},
\end{equation}
with projection threshold $\eta=0.5$ and $\betaH$ controlled by a deterministic continuation schedule in the
quantitative benchmark runs.
Design sensitivities $\partial c/\partial\rho_e$ are computed analytically and used
by the Optimality Criteria (OC) update rule~\cite{bendsoe2003topology} with
bisection to enforce the volume constraint at each SIMP iteration.
In the quantitative benchmark runs reported later, these nominal parameters are
embedded in a deterministic continuation schedule that ramps $\penal$, $\beta$,
$\rmin$, and the OC move limit; the exact schedule is stated in
Section~\ref{sec:results:setup}.
For reporting and validity checks, the grayness metric is defined as
\[
  g = \frac{4}{\nelem}\sum_{e=1}^{\nelem}\rho_e(1-\rho_e),
\]
so $g=0$ denotes a fully binary design and larger values indicate more
intermediate-density material.

\subsection{Matrix-Free \texorpdfstring{$\Kmat\mathbf{v}$}{Kv} Operator:
            Three-Stage Baseline}
\label{sec:method:baseline}

Explicit CSR assembly of $\Kmat$ is infeasible at the million-element scale on a
single GPU: for a trilinear hexahedral mesh each element
contributes $24\times 24$ entries indexed by 24 degrees of freedom (DOFs), so the
CSR index arrays alone occupy $\Ocal(576\,\nelem)$ integers---exceeding 24\,GB
\VRAM{} once explicit sparse assembly reaches the low-million-element regime on an
RTX\,4090.
Instead, the matrix--vector product $\mathbf{w} \leftarrow \Kmat\mathbf{v}$ is
evaluated element-by-element without ever forming $\Kmat$ globally:
\begin{equation}
  \mathbf{w} = \sum_{e=1}^{\nelem}
  \mathbf{B}_e^{\top}
  \bigl(\rho_{\min} + (1-\rho_{\min})\,\rho_e^{\penal}\bigr)\,
  \KE
  \mathbf{B}_e\,\mathbf{v},
  \label{eq:matfree}
\end{equation}
where $\mathbf{B}_e$ is the Boolean gather/scatter matrix from
\Cref{eq:simp}.
The baseline Python/CuPy implementation decomposes this into three sequential
kernel launches:

\begin{enumerate}[nosep,leftmargin=*,labelsep=0.6em,align=left,label=\textbf{Stage \arabic*.}]
  \item \textbf{Gather}: For each element $e$, copy
        $\mathbf{u}_e = \mathbf{u}[\mathrm{edof}_e] \in\mathbb{R}^{24}$ from the
        global displacement vector using the precomputed DOF index table
        $\mathrm{edof}\in\mathbb{Z}^{\nelem\times 24}$.
  \item \textbf{Batched GEMM}: Compute
        $\mathbf{f}_e =
        \bigl(\rho_{\min} + (1-\rho_{\min})\,\rho_e^{\penal}\bigr)\,
        \KE\,\mathbf{u}_e$
        using CuPy's batched matrix-multiplication path, which dispatches to
        cuBLAS over all $\nelem$ elements simultaneously.
  \item \textbf{Reduction scatter}: Accumulate
        $\mathbf{w}[\mathrm{edof}_e] \mathrel{+}= \mathbf{f}_e$
        with CuPy's histogram-style reduction back to the global output vector.
\end{enumerate}

This three-stage decomposition requires three full DRAM round-trips per
matrix--vector product: Gather reads $24\,\nelem$ doubles, batched GEMM reads the
element vectors again and writes results, and Scatter reads those results and
writes back to the global vector.
The intermediate per-element arrays---$\mathbf{u}_{\mathrm{elem}}$ and
$\mathbf{f}_{\mathrm{elem}}$---each occupy $24\,\nelem\times 8\,\text{bytes}$;
at $\nelem = 2\times 10^6$ these two buffers total 768\,MB, a non-trivial fraction
of the 24\,GB \VRAM{} budget.
More critically, these buffers are written by one kernel and read by the next,
forcing them to transit off-chip DRAM.
For the tested RTX\,4090, GDDR6X bandwidth peaks at 1.008\,TB/s; for reference,
the 80\,GB A100's HBM2e bandwidth is up to 1.94\,TB/s.
This off-chip traffic is what makes the baseline memory-bound at all sizes tested.

\subsection{Fused Gather--GEMM--Scatter Kernel}
\label{sec:method:fused}

The fused kernel eliminates the intermediate arrays by performing all three stages
within a single CUDA kernel launch, keeping $\mathbf{u}_e$ and $\mathbf{f}_e$ in
thread registers and shared memory throughout.
The implemented FP32 kernel uses one thread per finite element and 128 threads per
block.
The launch configuration is therefore
$(\lceil \nelem / 128 \rceil,\,1,\,1)$ blocks $\times$ $(128,\,1,\,1)$ threads,
with each thread gathering one element, applying the $24\times24$ dense multiply,
and atomically scattering its 24 contributions.

\begin{figure}[t]
  \centering
  \includegraphics[width=\linewidth]{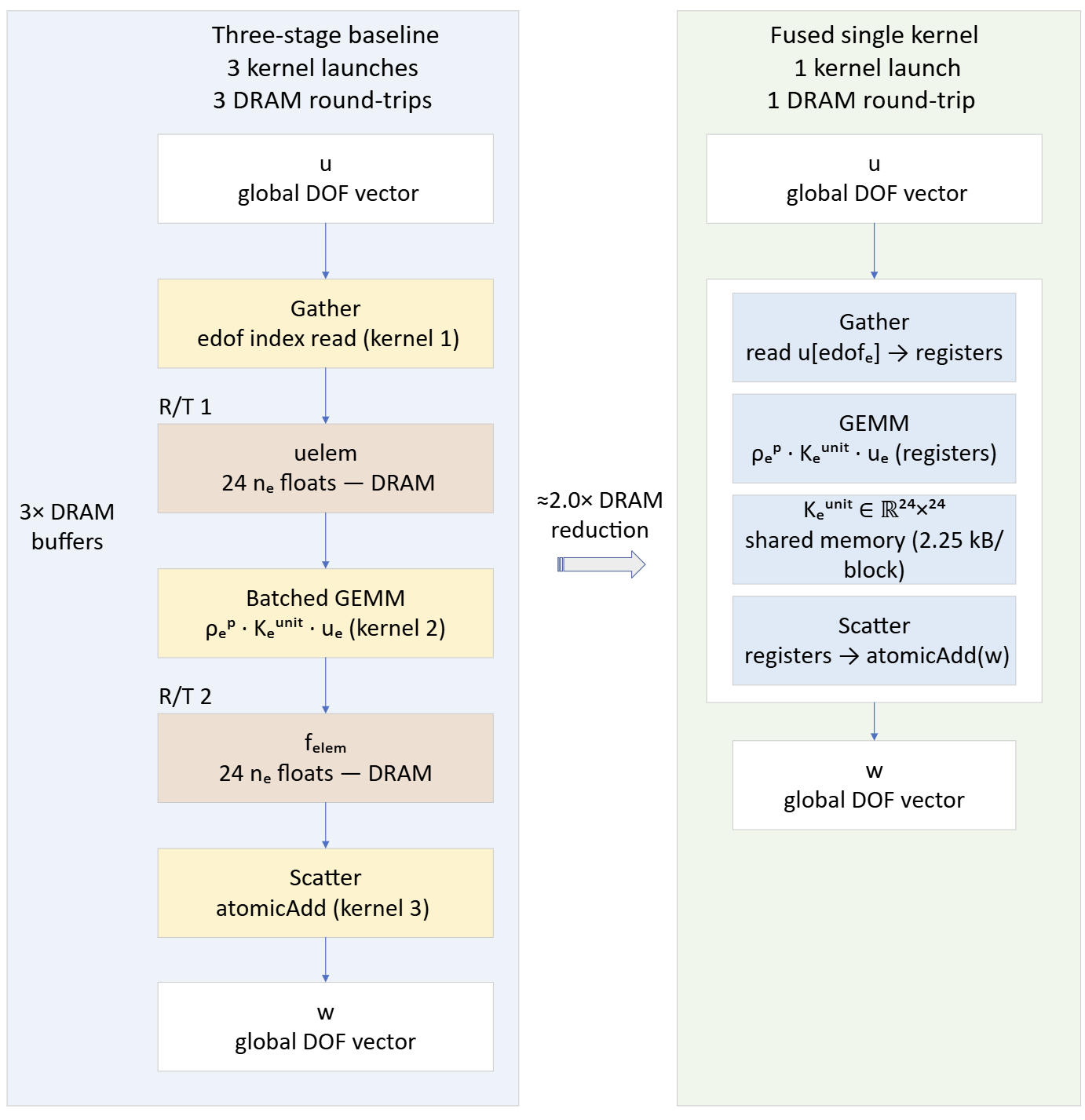}
  \caption{Comparison of the three-stage baseline pipeline (left) and the fused
           single-kernel implementation (right).
           The baseline requires three separate CUDA kernel launches per
           matrix--vector product, each forcing the intermediate per-element
           arrays $\mathbf{u}_{\mathrm{elem}}$ and $\mathbf{f}_{\mathrm{elem}}$
           to transit DRAM.
           The fused kernel keeps these in thread registers and shared memory,
           substantially reducing effective off-chip traffic
           (theoretical element-data reduction $\approx 2\times$).}
  \label{fig:pipeline}
\end{figure}

\begin{algorithm}[t]
\caption{Fused Gather--GEMM--Scatter kernel (per thread)}
\label{alg:fused}
\begin{algorithmic}[1]
\Require $\mathrm{edof}[\nelem,24]$, $\mathbf{v}[\ndof]$, $\KE[24,24]$,
         $\bm{\rho}[\nelem]$, $p$, $\rho_{\min}$
\Ensure  $\mathbf{w}[\ndof]$ (atomic-accumulated)
\State $e \leftarrow (\mathrm{block\ index})\cdot(\mathrm{block\ size}) + \mathrm{local\ thread\ index}$
\If{$e \ge \nelem$}
    \State \textbf{return}
\EndIf
\State \textbf{Shared memory}: cooperatively load $\KE[24,24]$ once per block
\State $k_e \leftarrow \rho_{\min} + (1-\rho_{\min})\,\rho_e^p$   \Comment{SIMP scaling}
\State Gather local DOFs: $u_j \leftarrow \mathbf{v}[\mathrm{edof}[e,j]]$ for $j=0,\ldots,23$
\For{$i = 0,\ldots,23$}
    \State $f_i \leftarrow 0$
    \For{$j = 0,\ldots,23$}
        \State $f_i \mathrel{+}= k_e \cdot \KE[i,j] \cdot u_j$
    \EndFor
    \State Atomically accumulate $f_i$ into $\mathbf{w}[\mathrm{edof}[e,i]]$
\EndFor
\end{algorithmic}
\end{algorithm}

\noindent\textbf{Shared memory layout.}
The unit element stiffness $\KE\in\mathbb{R}^{24\times 24}$ is the same for all
elements (it depends only on the element geometry and material constants, not on
$\rho_e$); it occupies $24^2\times 4\,\text{bytes} = 2.25\,\text{kB}$ in shared
memory and is loaded once per thread block via a cooperative broadcast before the
main loop.
Threads then read rows of $\KE$ from shared memory during the inner GEMM loop,
achieving a shared-memory bandwidth amplification of $24\times$ versus a naive
global-memory approach.

\noindent\textbf{Memory traffic analysis.}
Let $B_{\text{elem}}$ denote bytes per element per $\Kmat\mathbf{v}$ call.
The baseline three-stage path transfers:
\begin{multline}
  B_{\text{baseline}} = \underbrace{24\!\cdot\! 8}_{\text{gather out}}
    + \underbrace{24\!\cdot\! 8}_{\text{GEMM in (elem)}} \\
    + \underbrace{24\!\cdot\! 8}_{\text{GEMM out (elem)}}
    + \underbrace{24\!\cdot\! 8}_{\text{scatter in}}
    = 768\;\text{bytes/element}.
\end{multline}
The fused kernel eliminates the two intermediate passes; the minimum traffic is:
\begin{align}
  B_{\text{fused}} &= \underbrace{24\!\cdot\! 8}_{\text{gather global read}}
    + \underbrace{24\!\cdot\! 8}_{\text{scatter global write}}
    = 384\;\text{bytes/element},
\end{align}
a factor of $2\times$ reduction in element-data DRAM traffic.
Including one read of the $\mathrm{edof}$ table (96\,bytes/element) and the
$\rho$ array (4\,bytes/element), the theoretical minimum DRAM reads rise to
$\approx 480\,\text{bytes/element}$ for the fused kernel versus
$\approx 868\,\text{bytes/element}$ for the baseline, still a factor of $1.8\times$.
In practice, L2 cache captures repeated $\mathrm{edof}$ accesses and the runtime
speedups observed in Section~\ref{sec:results:profiling} are consistent with a
substantially larger reduction in effective off-chip traffic.

\noindent\textbf{CuPy runtime compilation.}
The fused kernel is written as a CUDA C string and compiled at first use through
CuPy's runtime kernel-compilation interface, which invokes NVRTC (NVIDIA Runtime Compilation)
without requiring a standalone CUDA toolchain.
The kernel string is approximately 80 lines of CUDA C, embedded directly in the
Python source file; no separate CUDA source file or offline compiler invocation is required.
This design preserves the fully Python-native workflow of the three-stage
baseline used throughout the paper while adding near-native CUDA performance.

\subsection{BF16 WMMA Tensor-Core Variant}
\label{sec:method:bf16}

Modern NVIDIA GPUs expose tensor cores through the WMMA (Warp-Level Matrix
Multiply--Accumulate) API, which computes $\mathbf{C} \leftarrow \mathbf{A}\mathbf{B} + \mathbf{C}$
for small fixed-size tile shapes in hardware-accelerated mixed-precision arithmetic.
On the RTX\,4090 (Ada Lovelace architecture) the supported BF16 WMMA shape is
$16\times 16 \times 16$ (m,n,k), accumulating into FP32.
BF16 represents values with a 7-bit mantissa and 8-bit exponent (same dynamic range
as FP32) and achieves $2\times$ the peak throughput of FP32 CUDA cores on the
RTX\,4090: 165.2\,TFLOP/s (BF16 tensor cores) versus 82.6\,TFLOP/s (FP32).

\begin{figure}[!htbp]
  \centering
  \includegraphics[width=\linewidth]{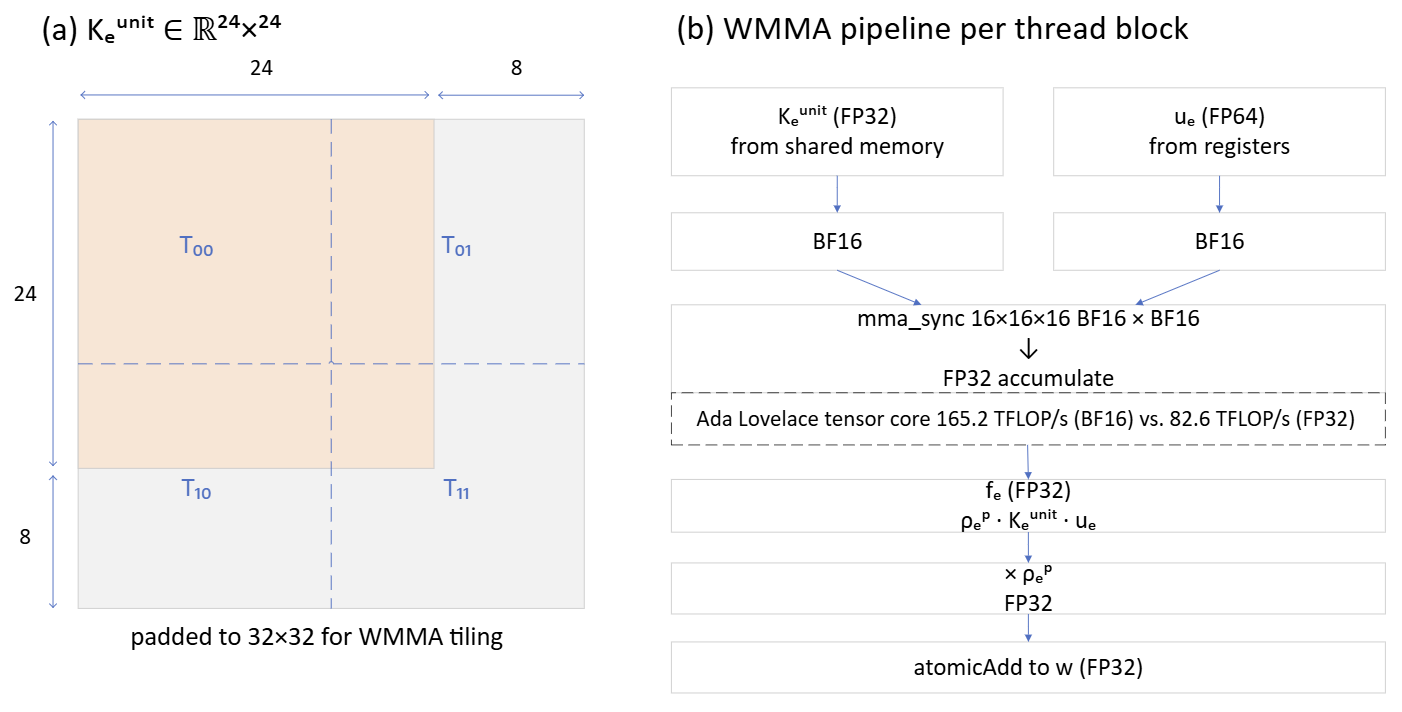}
  \caption{BF16 WMMA tensor-core tiling for the per-element stiffness multiply.
           \emph{Left}: the $24\times 24$ element stiffness matrix $\KE$ is
           zero-padded to $32\times 32$ and covered by four $16\times 16$
           output tiles, each accumulated from two $16\times 16\times 16$
           WMMA multiplies along the padded $k$ dimension.
\emph{Right}: 128 elements are batched per thread block (8 warps),
           with 16 elements processed per warp and 4 tensor-core matrix-multiply calls per
            warp.
            The WMMA multiply uses BF16 fragments; FP32 accumulation and the
            final atomic scatter maintain output fidelity.}
  \label{fig:bf16wmma}
\end{figure}

\noindent\textbf{Kernel design.}
The per-element GEMM $\mathbf{f}_e = k_e\,\KE\,\mathbf{u}_e$ involves a
$24\times 24$ matrix times a $24\times 1$ vector.
Directly mapping this to WMMA requires padding to multiples of 16:
$\KE$ is zero-padded to $32\times 32$, and $\mathbf{u}_e$ is padded to $32\times 16$
(one column per element with 15 zero columns).
Each CUDA thread block processes 128 elements, with 8 warps per block and
16 elements assigned to each warp.
The launch configuration is therefore $(\lceil\nelem/128\rceil,\,1,\,1)$ blocks
$\times$ $(256,\,1,\,1)$ threads.
Within each warp, 4 tensor-core matrix-multiply calls tile the $32\times 32$ output with
$16\times 16$ fragments, yielding the product $\KE\,[\mathbf{u}_{e_0}\,|\,\ldots\,|\,\mathbf{u}_{e_{15}}]$
simultaneously for all 16 elements in the block.
The SIMP scaling $k_e$ is applied post-WMMA in FP32, and the scatter uses
atomic accumulation with FP32 precision.

\noindent\textbf{Memory architecture implications.}
The BF16 WMMA path stores the gathered displacement values and the padded element
matrix in FP32 and casts them to BF16 immediately before loading WMMA fragments.
The SIMP penalty vector is maintained in FP32 throughout, and the scatter output is
written in FP32.
When this kernel is embedded in the mixed-precision solver, the resulting
matvec is used inside an FP32 CG loop; in the BF16 experiments reported later, an FP32
outer residual-correction loop wraps that inner BF16 solve.
This FP32$\to$BF16$\to$FP32 mixed-precision cascade is the defining characteristic of the
BF16 integration path and the source of the convergence difficulties analyzed in
the next section.

\subsection{BF16 Arithmetic in the Conjugate Gradient Solver:
            Convergence Analysis}
\label{sec:method:bf16conv}

\noindent\textbf{Residual-correction formulation used in the reported experiments.}
The reported BF16 experiments use a practical two-precision residual-correction
loop inspired by iterative refinement~\cite{carson2018threeprecision}:
\begin{align}
  \mathbf{r}^{(k)} &= \mathbf{f} - \Kmat\mathbf{u}^{(k)} \quad (\text{FP32}), \\
  \mathbf{e}^{(k)} &\approx \Kmat^{-1}\mathbf{r}^{(k)}  \quad (\text{BF16 inner CG}), \\
  \mathbf{u}^{(k+1)} &= \mathbf{u}^{(k)} + \mathbf{e}^{(k)}.
\end{align}
The classical convergence lens is still the standard
$\eps_{\ell}\cdot\kappa(\Kmat) \leq c$ for a modest constant $c < 1$ and
working-precision unit roundoff $\eps_{\ell}$~\cite{carson2018threeprecision}.
For BF16, we use the Carson--Higham unit-roundoff convention
$\eps_{\mathrm{BF16}} = 2^{-8} \approx 3.9\times 10^{-3}$ throughout.

\noindent\textbf{Condition number of the SIMP stiffness matrix.}
The condition number $\kappa(\Kmat)$ grows with both the SIMP penalty exponent
and the density contrast between solid and void elements.
For a well-converged SIMP design with large penalization and near-void regions,
the modulus contrast between solid ($\rho_e = 1$) and near-void ($\rho_e \approx \rho_{\min}$)
elements grows as $1/\rho_{\min}^{\penal}$, driving $\kappa(\Kmat)$ well
above the BF16 threshold at moderate penalization.
A companion condition-number-estimation study estimates $\kappa(\Kmat)$ via matrix-free power iteration
for the largest eigenvalue and inverse iteration for the smallest, evaluating the
uniform-density initialization states ($\rho=0.5$) across three mesh
sizes (64\,k, 216\,k, and 512\,k) and two penalization levels
($\penal \in \{3.0, 5.0\}$).
No explicit $\Kmat$ is assembled; both extremal eigenvalues are estimated
entirely through matrix-free $\Kmat\mathbf{v}$ applications
(Algorithm~\ref{alg:fused}), making the procedure available at all tested sizes.
Power iteration targets a relative change tolerance of $10^{-6}$ on successive
eigenvalue estimates, and inverse iteration uses a $10^{-4}$ relative change
tolerance on the reciprocal Rayleigh estimate; the supplementary tabulation
records the realized outer-iteration counts for each row.
The current workflow uses fixed random start vectors (seed 42 for power
iteration and seed 123 for inverse iteration); each inverse-iteration inner
solve uses SciPy CG with relative and absolute tolerance $10^{-8}$ and a
2{,}000-iteration cap.
In the reported rows, the power iteration reaches the
50-step cap before meeting the $10^{-6}$ target, so the reported
$\lambda_{\max}$ and $\kappa$ values should be read as conservative estimates.
The reported estimates indicate that $\kappa(\Kmat)$ exceeds
$1/\varepsilon_{\mathrm{BF16}} \approx 256$ for the benchmark systems studied
here (see Section~\ref{sec:results:kappa}).

\noindent\textbf{Convergence barrier.}
The BF16 convergence condition requires:
\begin{equation}
  \eps_{\mathrm{BF16}} \cdot \kappa(\Kmat) < 1
  \;\iff\;
  \kappa(\Kmat) < \frac{1}{\eps_{\mathrm{BF16}}} \approx 256.
  \label{eq:bf16barrier}
\end{equation}
Direct power-iteration estimates (Section~\ref{sec:results:kappa}) indicate that
the tested systems lie firmly and deeply in the non-convergence zone:
$\kappa(\Kmat) \approx 6.1\times10^{5}$ at 64\,k, $\approx 1.3\times10^{6}$
at 216\,k, and $\approx 2.3\times10^{6}$ at 512\,k elements, giving
$\varepsilon_{\mathrm{BF16}}\cdot\kappa \approx 2.4\times10^{3}$--$9.1\times10^{3}$---more than an
order of magnitude above the threshold across all tested sizes.
The BF16 residual-correction experiments in
Section~\ref{sec:results:bf16} stagnate at large compliance error precisely
because the inner BF16 solve is unable to reduce the outer FP32 residual
below the BF16 noise floor in this regime.

\noindent\textbf{Path forward.}
The BF16 GEMM-stage throughput advantage (the separate $14.3\times$ PyTorch BF16
GEMM proxy timing at 512\,k elements discussed in Section~\ref{sec:results:profiling}) motivates
using the BF16 kernel as a multigrid smoother rather than as a preconditioner for
the full-system CG.
In a V-cycle smoother, the spectrum of the residual presented to each level is
bounded by the coarse-grid correction, so the effective $\kappa$ seen by the BF16
smoother can be kept below the $1/\eps_{\mathrm{BF16}} \approx 256$ threshold.
This direction is identified as the primary future extension in
Section~\ref{sec:discussion}.

\subsection{Preconditioned Conjugate Gradient and Warm-Start Strategy}
\label{sec:method:pcg}

The global linear system $\Kmat\mathbf{u} = \mathbf{f}$ is solved at each SIMP
iteration using Jacobi-preconditioned conjugate gradient (PCG)~\cite{hestenes1952cg,saad2003iterative}.
The Jacobi preconditioner $\mathbf{M} = \mathrm{diag}(\Kmat)$ is formed exactly
in $\Ocal(\nelem)$ via a secondary matrix-free pass that accumulates diagonal
contributions element-by-element, without any explicit assembly.
Despite its simplicity, Jacobi preconditioning requires
$\Ocal(\sqrt{\kappa(\Kmat)})$ CG iterations for convergence;
in practice this yields a few hundred iterations per SIMP step for the cantilever
problem, while the torsion benchmark often approaches the 1,000-iteration cap.
The Jacobi PCG used in the core FP64/FP32/fused solver path admits a pure
Python/CuPy implementation with no standalone CUDA build step.
The separate tensor-core GEMM proxy benchmark reported later uses PyTorch only
for timing those BF16/FP16 GEMM proxy calls; it is not part of the core solver path.

\noindent\textbf{Warm-start.}
Successive SIMP iterations produce design fields that evolve slowly once past the
first 10--20 iterations; the displacement solution $\mathbf{u}^{(k)}$ from
iteration $k$ is therefore used as the initial guess for iteration $k+1$.
Warm-start is enabled in all quantitative SIMP benchmarks and materially reduces
the later-iteration CG workload relative to repeated cold starts.

\noindent\textbf{Convergence criterion.}
In the reported implementation, CG is run until the unpreconditioned relative
residual satisfies $\|\mathbf{r}_k\|/\|\mathbf{f}\| \leq 10^{-5}$, or a maximum of
1,000 iterations is reached.
This is the stopping rule used throughout the reported experiments and across the
FP32/BF16 matrix-free solver variants.

\subsection{Roofline Analysis}
\label{sec:method:roofline}

The roofline model~\cite{williams2009roofline,ding2019gpuroofline} bounds kernel throughput by the
minimum of compute peak ($\Pi$, in FLOP/s) and memory bandwidth times arithmetic
intensity ($I$, in FLOP/B):
$P \leq \min(\Pi,\; I \cdot b)$,
where $b$ is the memory bandwidth ceiling.
Table~\ref{tab:roofline} summarizes the RTX\,4090 performance ceilings and the
arithmetic intensity of the fused and BF16 kernels.

\begin{table}[h]
\centering
\caption{RTX\,4090 roofline ceilings and idealized arithmetic intensity
         ($I_{\text{ideal}}$) for the full-kernel variants reported in the paper~\cite{nvidiaada2023}.
         Dashes indicate hardware-ceiling rows for which arithmetic intensity is
         not applicable. All three full-kernel variants remain
         DRAM-bandwidth-bound on the tested RTX\,4090.
         Section~\ref{sec:results:profiling} separately reports
         profiling-traffic arithmetic intensity $I_{\text{profile}}$, which
         includes index and density reads.}
\label{tab:roofline}
\resizebox{\linewidth}{!}{%
\begin{tabular}{lrrrr}
\toprule
Variant &
  $\Pi$ (TFLOP/s) &
  $b$ (TB/s) &
  Ridge (FLOP/B) &
  $I_{\text{ideal}}$ (FLOP/B) \\
\midrule
FP64 CUDA cores   & 1.29  & 1.008 & 1.29  & --- \\
FP32 CUDA cores   & 82.6  & 1.008 & 81.9  & --- \\
BF16 tensor cores & 165.2 & 1.008 & 163.9 & --- \\
\midrule
Baseline 3-stage (FP64)  & ---   & 1.008 & ---  & $\approx 3.2$ \\
Fused FP32 kernel        & ---   & 1.008 & ---  & $\approx 5.8$ \\
Fused BF16 full kernel   & ---   & 1.008 & ---  & $\approx 5.8$ \\
\bottomrule
\end{tabular}
}%
\end{table}

\noindent
The idealized arithmetic intensity of the fused FP32 kernel is estimated as
\[
  I_{\text{ideal,fused}}
    = \frac{2\times 24^2\;\text{FLOP}}{2\times 24\times 4\;\text{bytes}}
    \approx 6\;\text{FLOP/B},
\]
and a coarse implementation-level accounting gives an effective
$I_{\text{ideal}} \approx 5.8\,\text{FLOP/B}$---well below the
FP32 ridge point of $81.9\,\text{FLOP/B}$.
The fused kernel is therefore DRAM-bandwidth-bound across all tested sizes, and
the roofline-level speedup over the three-stage baseline is explained by lower
DRAM traffic rather than by higher arithmetic intensity.
Under the profiling traffic model that also counts $\mathrm{edof}$ index reads
and per-element density reads, the reported values are
$I_{\text{profile}}=1.33$ (FP64 three-stage) and $3.95$ (fused FP32) in
\Cref{tab:bw_utilization}.
The reported BF16-versus-FP64 $14.3\times$ GEMM proxy timing at 512\,k elements
shows that the isolated GEMM sub-operation can be accelerated dramatically by
BF16 tensor-core arithmetic in a separate PyTorch proxy benchmark.
The paper does not report a separate stage-level roofline derivation for that
GEMM proxy.
The only full-kernel roofline points used in the paper are those listed in
\Cref{tab:roofline}, and under that accounting the full fused-BF16 kernel
remains bandwidth-bound because gather and scatter preserve the same irregular
global-memory traffic as the fused FP32 path.

\subsection{Software Architecture}
\label{sec:method:software}

The complete solver is implemented in Python\,3.11 and CuPy\,13 and runs on a
single consumer GPU (NVIDIA RTX\,4090, 24\,GB GDDR6X) without any standalone CUDA
build step.
The software layers are:

\begin{enumerate}[nosep]
  \item \textbf{Problem setup}: mesh generation, DOF table ($\mathrm{edof}$)
        construction, boundary condition assembly, and filter kernel
        construction---all in NumPy/CuPy.
  \item \textbf{Fused CUDA kernel}: the CUDA C string is compiled once at first use
        through CuPy's runtime kernel-compilation interface and cached; subsequent calls reuse the
        compiled binary.
  \item \textbf{PCG solver loop}: pure Python with CuPy array operations;
        the fused kernel is called as a function object for the $\Kmat\mathbf{v}$
        product.
  \item \textbf{SIMP outer loop}: OC update, filter application, Heaviside
        projection, and convergence check in NumPy/CuPy.
  \item \textbf{Results I/O}: density fields and benchmark metrics are serialized
        to NumPy array and CSV formats for post-processing.
\end{enumerate}

All data remain on the GPU throughout the inner PCG loop; only scalar convergence
monitors and per-SIMP-iteration compliance values are transferred to the CPU
through ordinary scalar conversion.
In the reported SIMP scaling runs, the fused path uses about 5.0\,GB at
$\nelem \approx 2\times 10^6$ and 10.56\,GB for an 8\,M-element FEA-only solve,
leaving substantial headroom within the 24\,GB VRAM limit.

\section{Experimental Results}
\label{sec:results}

\subsection{Experimental Setup}
\label{sec:results:setup}

\begin{sloppypar}
\noindent\textbf{Hardware.}
All GPU experiments are conducted on a single NVIDIA RTX\,4090~\cite{nvidiaada2023}
(Ada Lovelace; 24\,GB GDDR6X; 1.008\,TB/s bandwidth;
165.2\,/\,82.6\,/\,1.29\,TFLOP/s for BF16/FP32/FP64 respectively).
The reported results were generated on Microsoft Windows 10.0.26200.8037 with
NVIDIA driver 595.71, in Python\,3.11 under CuPy\,13.6.0,
PyTorch\,2.5.1 built against CUDA\,12.1, NumPy\,2.2.6, SciPy\,1.15.3,
Matplotlib\,3.10.7, scikit-image\,0.25.2, Pandas\,2.3.3, and PyVista\,0.46.3;
the CuPy runtime reports CUDA runtime version 12090, and the board uses the
default 450\,W power limit.
This CUDA\,12.1/12.9 split reflects PyTorch's bundled CUDA\,12.1 user-space
runtime alongside the installed CUDA\,12.9 runtime used by CuPy; that mixed
configuration is supported by the installed NVIDIA driver on this host.
The core solver path uses CuPy; PyTorch is used only in the hot-path microbenchmark
script for the separate BF16/FP16 GEMM proxy timings, while the render scripts
additionally use PyVista and scikit-image.
\end{sloppypar}

\noindent\textbf{Benchmark problems.}
Two main quantitative benchmark families are used throughout (cantilever and
torsion), together with two 187{,}500-element hard-problem stress tests (MBB
and bridge):

\begin{itemize}[nosep]
  \item \textbf{Cantilever beam}: domain $2\times1\times0.5$ with fixed left face
        and a unit downward point load at the right-face midpoint.
        All cantilever runs use $\vf = 0.30$ and an initial filter radius
        $\rmin = 1.5$.
        The hot-path microbenchmark covers 64\,k, 216\,k, 512\,k, and 1\,M
        elements. The canonical cold-start FEA scaling figure covers 216\,k,
        512\,k, 1\,M, 2.0\,M, 4.9\,M, and 8\,M elements, while the end-to-end
        SIMP scaling study reports 216\,k, 512\,k, 1\,M, 2.0\,M, and 4.9\,M
        elements.
  \item \textbf{MBB hard-problem stress test} ($150\times 50\times 25 =
        187{,}500$ elements; domain $3\times1\times0.5$; $u_x$ constrained
        along the left edge, $u_y$ constrained at $(3,0,0.25)$, and a unit
        downward point load applied at $(0,1,0.25)$; $\vf = 0.50$; initial
        $\rmin = 1.5$).
  \item \textbf{Bridge hard-problem stress test} ($150\times 50\times 25 =
        187{,}500$ elements; domain $3\times1\times0.5$; fixed support points
        on the left lower edge, roller-$x$ support points on the right lower
        edge, and a distributed downward load on the top edge; $\vf = 0.30$;
        initial $\rmin = 1.5$).
  \item \textbf{Torsion shaft} ($165\times 55\times 55 = 499{,}125$ elements;
        fixed left face, equal-and-opposite torsional loads on the top and bottom
        edges of the right face; $\vf = 0.25$; initial $\rmin = 1.5$).
\end{itemize}

An additional left-clamped, right-roller beam example is shown later as a
qualitative render companion to the bridge load family; it is not the source of
the bridge timing table.

The main cantilever and torsion optimization studies use 120 outer iterations
and the same fixed four-phase continuation schedule: iterations 1--15 use
$(\penal,\beta,m)=(1.5,1.0,0.20)$; iterations 16--40 use $(3.5,4.0,0.15)$
while reducing $\rmin$ toward 1.35; iterations 41--65 use
$(4.5,16.0,0.08)$ while reducing $\rmin$ toward 1.25; and iterations 66--120
use $(4.5,32.0,0.05)$ while reducing $\rmin$ toward 1.20, where $m$ is the OC
move limit.
If compliance rises above $1.12\times$ the current selected compliance, the run
restarts from the currently selected design field.
Table~\ref{tab:bridge_hard} is intentionally separate from this main protocol:
its truncated hard-problem stress test uses 60 outer iterations, cold-start
initialization, and no warm-start.
For the main SIMP-120 studies, reported compliance values and rendered
topologies therefore correspond to the \emph{selected design iterate} from each
run: the lowest-compliance iterate that passes the run's validity checks, not
necessarily the final iterate.
Appendix~\ref{app:provenance} defines this selected-versus-final convention
formally for all reader-facing tables and figures; the hard-problem table uses
the same best-valid selection rule, but on the shorter 60-step cold-start
schedule stated in its caption.
In the present implementation, an iterate becomes eligible for this selection
only once $\penal \ge 3.0$ and the grayness metric satisfies $g < 0.25$.
Here the grayness metric is
\[
  g = \frac{4}{\nelem}\sum_{e=1}^{\nelem}\rho_e(1-\rho_e),
\]
so $g=0$ denotes a fully binary design and larger values indicate more
intermediate-density material.
All SIMP runs use deterministic uniform-density initialization, so no random
seed is required for the optimization path itself.

\noindent\textbf{Solver variants.}
Three implementations are benchmarked:
\textsc{fp64} (three-stage baseline: gather + CuPy-dispatched batched DGEMM + histogram-style scatter reduction, FP64 throughout),
\textsc{fp32} (same three-stage pipeline in FP32 state, with the histogram-style scatter reduction accumulated in float64 before cast-back),
and \textsc{fused} (the single fused gather--GEMM--scatter kernel in FP32).
A fourth variant \textsc{bf16-wmma} is benchmarked for full fused-kernel matvec
throughput, for the separate BF16 GEMM proxy timing reported in
\Cref{tab:profiling}, and for representative mixed-precision linear-solve
experiments.
Warm-start is enabled for the main cantilever and torsion SIMP-120 benchmarks
(Section~\ref{sec:method:pcg}); the hard-problem SIMP-60 rows in
Table~\ref{tab:bridge_hard} are explicit cold-start runs without warm-start.
Timing is measured with CUDA events (device-side) for kernel profiling and
wall-clock time for end-to-end SIMP benchmarks.
The end-to-end SIMP wall times in \Cref{tab:simp_scaling,tab:torsion} are
single representative runs rather than sample means;
Tables~\ref{tab:repeat_study}--\ref{tab:highcap_validation} report the separate
repeat, determinism, and high-cap validation studies, and the bridge
hard-problem table states its mixed mean/single-run reporting explicitly in the
caption.

\FloatBarrier
\subsection{Hot-Path Microbenchmark and Single-Solve Scaling}
\label{sec:results:profiling}

\begin{figure}[!htbp]
  \centering
  \includegraphics[width=0.85\linewidth]{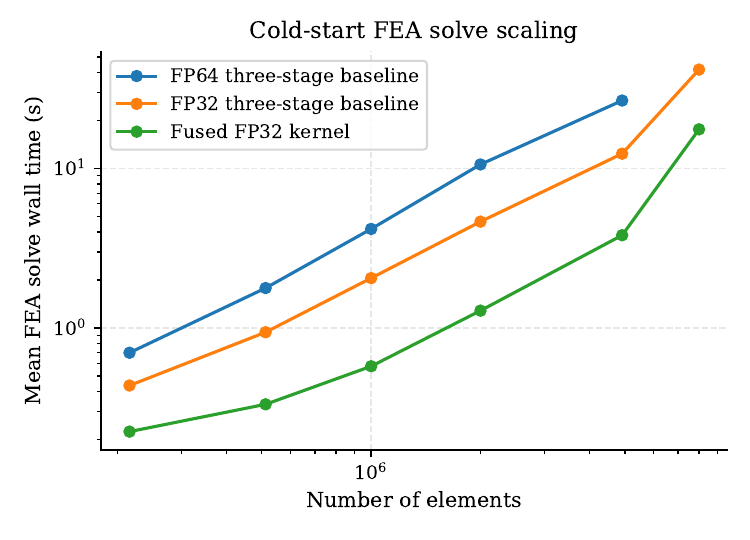}
  \caption{Cold-start FEA solve wall time versus element count for the three
             cantilever solver variants (RTX\,4090, log--log scale) at uniform
             density $\rho=0.5$ for $216{,}000$ to $8{,}000{,}000$ elements.
              Each point is a full Jacobi-PCG solve, not a single matvec timing.
              The 2\,M, 4.9\,M, and 8\,M points reach the current 1{,}000-iteration
              cap and should be read as capped stress-test timings rather than as
             fully converged solves.
             The 8\,M point is available only for the FP32 and fused paths in the
             current study.
             The fused path preserves the same linear-solve algorithm while reducing
             the cost of each operator application.}
  \label{fig:fea_scaling}
\end{figure}

\Cref{tab:profiling} reports a synthetic hot-path microbenchmark for the
gather/GEMM/scatter operator shape at four cantilever mesh sizes.
        The source script seeds random synthetic data with the same tensor shapes
        as the real operator, including random DOF-index patterns rather than the
        structured cantilever connectivity, using a fixed seed of 42 and an
        adaptive repeat count that targets about 50 ms of total measurement per
        size.
Figure~\ref{fig:fea_scaling} then reports cold-start wall-clock times for a
full FEA solve at uniform density $\rho=0.5$, providing a size-scaling check of
the cold-start linear-solve cost.

\begin{table}[H]
\centering
\caption{Synthetic hot-path timing breakdown ($\mu$s) on the RTX\,4090.
         $t_{\mathrm{full64}}$: three-stage FP64 pipeline (gather + batched DGEMM + scatter).
         $t_{\mathrm{fused}}$: single fused FP32 kernel.
         $t_{\mathrm{bf16}}$: single fused BF16\,WMMA kernel.
         Speedup is relative to $t_{\mathrm{full64}}$.}
\label{tab:profiling}
\resizebox{\linewidth}{!}{%
\begin{tabular}{lrrrrrrrr}
\toprule
Size ($\nelem$) &
  $t_{\mathrm{gather}}$ &
  $t_{\mathrm{DGEMM64}}$ &
  $t_{\mathrm{scatter}}$ &
  $t_{\mathrm{full64}}$ &
  $t_{\mathrm{fused}}$ &
  $t_{\mathrm{bf16}}$ &
  \makecell{Fused\\speedup} &
  \makecell{BF16\\speedup} \\
\midrule
64\,k  &  28.6 &  140.8 & 167.1 &  342.6 &  56.9 &  58.3 & 6.0$\times$ & 5.9$\times$ \\
216\,k &  70.8 &  452.3 & 315.6 & 1052.2 & 168.3 & 166.3 & 6.3$\times$ & 6.3$\times$ \\
512\,k & 183.4 & 1154.3 & 731.2 & 2409.0 & 364.1 & 361.9 & 6.6$\times$ & 6.7$\times$ \\
1\,M   & 351.3 & 2280.7 & 1423.2& 4615.8 & 765.3 & 739.0 & 6.0$\times$ & 6.2$\times$ \\
\bottomrule
\end{tabular}
}%
\end{table}

\begin{figure}[!htbp]
  \centering
  \includegraphics[width=0.85\linewidth]{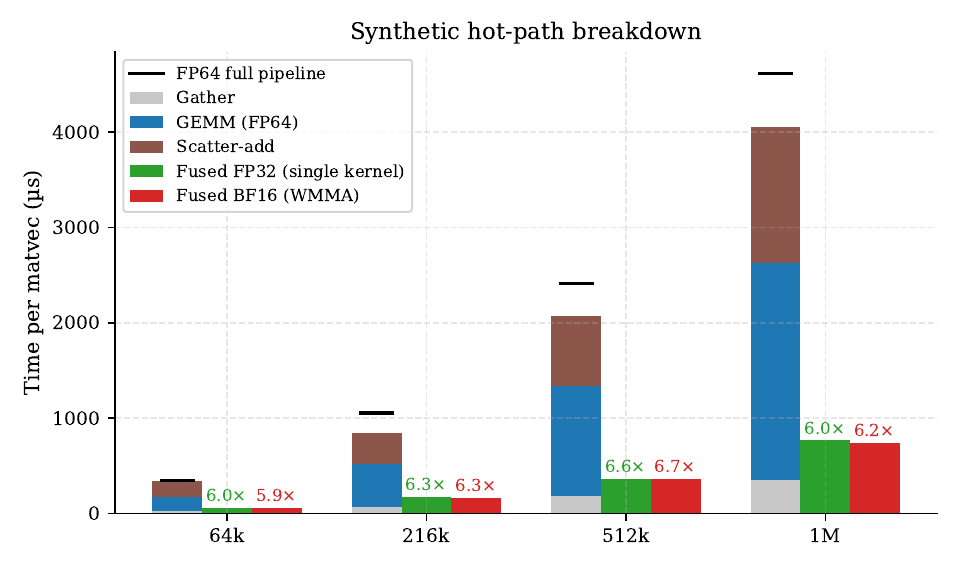}
  \caption{Synthetic hot-path timing breakdown at four element counts.
           Bars show the FP64 gather, batched DGEMM, and scatter components from
            the three-stage microbenchmark, together with the corresponding fused
            FP32 and fused BF16 kernels measured on the same synthetic input shapes.
            This figure illustrates operator-level cost composition; it is not a
            full SIMP or full FEA timing plot. The stacked gather/DGEMM/scatter
            bars represent the sum of separately measured stage durations, while
            the black markers indicate the full FP64 pipeline time from
            Table~\ref{tab:profiling}; the stage sum is therefore a lower bound on
            the complete FP64 path. Its gather/scatter locality
            should be read as a synthetic stress test rather than as a literal
             replay of the structured cantilever connectivity pattern.}
  \label{fig:profiler}
\end{figure}

In the synthetic operator microbenchmark, the fused kernel achieves a
\textbf{6.0--6.6$\times$} speedup over the FP64 three-stage baseline across the
full size range.
Profiling-guided traffic accounting indicates that the fused kernel reduces
effective off-chip traffic substantially,
consistent with the theoretical analysis in Section~\ref{sec:method:fused}.
The remaining gap between the theoretical $2\times$ element-data reduction
(or roughly $1.8\times$ once index and density reads are included) and the
observed $6\times$ speedup is consistent with two additional effects: (i)
elimination of kernel-launch overhead across the three GPU-resident stages and
(ii) improved warp utilization from the regular 128-thread block configuration
of the fused FP32 kernel. Because
the stage timings in \Cref{tab:profiling} are measured independently, the stacked
component sum in \Cref{fig:profiler} is a lower bound on the full FP64 pipeline
time rather than a complete decomposition of every pipeline overhead.

\noindent\textbf{GEMM-stage BF16 proxy timing.}
While the \emph{total} per-matvec speedup of the BF16 WMMA kernel is comparable to
the fused FP32 kernel (both around 6$\times$ vs.\ the FP64 pipeline), the GEMM
proxy benchmark reported by the separate profiling study shows dramatic
improvement: at 512\,k elements, the BF16 GEMM proxy
takes 80.6\,$\mu$s versus 1154.3\,$\mu$s for the FP64 batched DGEMM---a
\textbf{14.3$\times$ BF16-versus-FP64 GEMM proxy speedup}.
This proxy is measured separately with PyTorch BF16 GEMM on the same tensor
shapes; it is not an instrumented stage timing extracted from the custom CuPy
WMMA kernel itself.
The reason the full-pipeline speedup saturates near 6$\times$ is that the
        gather and scatter stages (which still stage through the global-vector and
        reduction paths)
        become the dominant cost once the GEMM is accelerated by tensor cores:
at 512\,k elements, gather+scatter consumes 914.6\,$\mu$s of the FP64 pipeline
but only about 281.3\,$\mu$s of non-GEMM work in the fused BF16 path
($361.9 - 80.6 \approx 281.3$)\footnote{Computed from the canonical 512\,k profiler row:
full fused BF16 kernel minus the separate BF16 GEMM proxy timing. This subtraction
is a useful upper-bound accounting argument, not an instrumented stage timing inside
the CuPy WMMA kernel.}---so
even with zero-cost GEMM, the achievable speedup would be capped at
$2409/281.3 \approx 8.6\times$.
This analysis motivates the future direction of a fully fused BF16 kernel that
also eliminates the FP64 gather--scatter round-trip.

\noindent\textbf{Bandwidth utilization on the actual operator.}
To complement the synthetic microbenchmark, a separate CUDA-event measurement
on the actual cantilever operator (50 timed iterations with 10 warm-up
iterations) yields
effective memory bandwidths of 61--179\,GB/s (6--18\% of the RTX\,4090's
1,008\,GB/s peak DRAM bandwidth) for the fused FP32 kernel, versus
20--43\,GB/s (2--4\% of peak) for the FP64 three-stage baseline
(\Cref{tab:bw_utilization}).
Both paths sit far below the roofline ridge point of
$\approx82$\,FLOP/B (82.6\,TFLOP/s $\div$ 1,008\,GB/s), indicating
that the gather--GEMM--scatter operator is memory-bandwidth limited
regardless of precision.
The isolated per-call speedup on the actual cantilever operator is
\textbf{8.9--13.8$\times$}---higher than the 6.0--6.6$\times$ from the
synthetic benchmark---because the actual Python/CuPy dispatch overhead
is also eliminated: the fused kernel replaces three separate CuPy API
calls (one per stage) with a single runtime-compiled kernel invocation.

\begin{table}[H]
\centering
\caption{Memory bandwidth utilization of the actual cantilever operator,
         measured with CUDA events on the RTX\,4090 (50 timed iterations,
         10 warm-up).  $I_{\text{profile}}$ = profiling-traffic arithmetic intensity
         (FLOP/byte), including index and density reads.
         Effective bandwidth is theoretical bytes moved divided by
         measured wall time under the profiling traffic model.}
\label{tab:bw_utilization}
\resizebox{0.90\linewidth}{!}{%
\begin{tabular}{lrrrrrrr}
\toprule
Size & \multicolumn{3}{c}{FP64 three-stage} & \multicolumn{3}{c}{Fused FP32} & Speedup \\
\cmidrule(lr){2-4}\cmidrule(lr){5-7}
     & $t$ ($\mu$s) & BW (GB/s) & $I_{\text{profile}}$ & $t$ ($\mu$s) & BW (GB/s) & $I_{\text{profile}}$ & \\
\midrule
64\,k  & 2736 & 20.3 & 1.33 & 307  & 61.0  & 3.95 & 8.9$\times$ \\
216\,k & 4871 & 38.5 & 1.33 & 353  & 178.8 & 3.95 & 13.8$\times$ \\
512\,k & 11291 & 39.4 & 1.33 & 833  & 179.4 & 3.95 & 13.6$\times$ \\
1\,M   & 20285 & 42.8 & 1.33 & 1700 & 171.8 & 3.95 & 11.9$\times$ \\
\bottomrule
\end{tabular}
}%
\end{table}

\FloatBarrier
\subsection{End-to-End SIMP Scaling: Cantilever Benchmark}
\label{sec:results:scaling}

\Cref{tab:simp_scaling} reports full SIMP-120 wall times and compliance values for
the cantilever benchmark across five mesh sizes.
The fused FP32 kernel achieves consistent speedups of \textbf{4.6--7.3$\times$}
over the FP64 three-stage baseline end-to-end.
Against the same-precision FP32 three-stage baseline, the same rows correspond
to 2.3--4.6$\times$ speedups.
These rows are representative single-run measurements from the dedicated scaling
workflow; Section~\ref{sec:results:stats} reports a separate five-repeat study at
216\,k and 512\,k to quantify run-to-run variability.

\begin{table}[H]
\centering
\caption{End-to-end SIMP-120 results on the cantilever benchmark (RTX\,4090).
         FP64 and FP32 denote the three-stage baselines.
         Reported compliance values are the selected compliances from the
          120-step continuation schedule.
         Speedup is FP64/Fused.
         These rows are representative single-run timings from the scaling
         workflow: each deposited (size, path) pair appears once in that
         workflow, while repeat variability at 216\,k and 512\,k is reported
         separately in Table~\ref{tab:repeat_study}. The 2\,M fused compliance
         deviates by 2.46\% from FP64; see Section~\ref{sec:discussion}.}
\label{tab:simp_scaling}
\begin{tabular}{lrrrrrrrr}
\toprule
\multirow{2}{*}{Size} &
\multirow{2}{*}{$\nelem$} &
\multicolumn{2}{c}{FP64} &
\multicolumn{2}{c}{FP32} &
\multicolumn{3}{c}{Fused FP32} \\
\cmidrule(lr){3-4}\cmidrule(lr){5-6}\cmidrule(lr){7-9}
 & & $t$ (s) & $c$ & $t$ (s) & $c$ & $t$ (s) & $c$ & Speedup \\
\midrule
216\,k &  216{,}000 &  80.4 & 2.172 &  80.5 & 2.174 &  17.5 & 2.174 & 4.6$\times$ \\
512\,k &  512{,}000 & 137.9 & 1.825 &  57.9 & 1.828 &  24.8 & 1.828 & 5.6$\times$ \\
1\,M   & 1{,}000{,}000 & 226.2 & 1.606 & 120.5 & 1.608 &  45.4 & 1.608 & 5.0$\times$ \\
2\,M   & 2{,}000{,}376 & 2610.2 & 1.421 & 1190.4 & 1.407 & 358.0 & 1.386 & 7.3$\times$ \\
4.9\,M & 4{,}913{,}000 & 6140.9 & 1.046 & 3000.6 & 1.040 & 996.9 & 1.045 & 6.2$\times$ \\
\bottomrule
\end{tabular}
\end{table}

Several observations merit discussion.
First, the FP32 path shows negligible difference relative to FP64 at 216\,k elements
(80.5\,s vs.\ 80.4\,s) but achieves 2.4$\times$ at 512\,k and 1.9$\times$ at 1\,M.
This is consistent with a bandwidth-bound workload in which FP32 halves the
per-entry data footprint relative to FP64.
At 216\,k, the non-FEA SIMP overhead is still large enough to dilute that
matvec-level advantage in the full end-to-end wall time.
Second, the fused kernel shows a \emph{higher} relative speedup at 2\,M (7.3$\times$)
than at 1\,M (5.0$\times$), for which one plausible explanation is cache behavior:
the $\Ke^{\mathrm{unit}}$ matrix (2.25\,kB shared memory) is reused across all
2\,M thread blocks, but at smaller sizes the per-block overhead of loading the
kernel launch context dominates.
We do not have cache-counter measurements for this interpretation, so
this should be read as a hypothesis rather than as a verified mechanism.
Third, the selected compliance values remain within about 0.2\% of the FP64
reference through 1\,M elements.
At 2\,M, the fused path deviates by 2.46\% and the FP32 path by 0.97\%, so the
exact values are reported explicitly rather than summarized as decimal-place parity.

\begin{figure}[!htbp]
  \centering
  \includegraphics[width=0.85\linewidth]{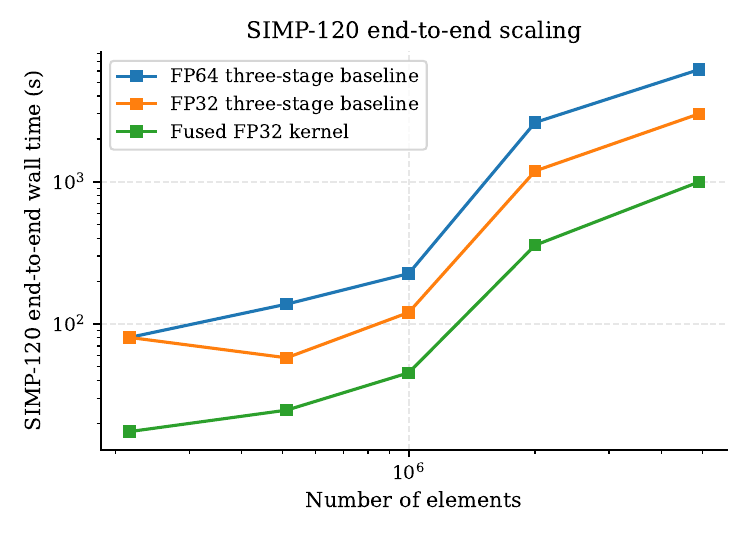}
\caption{End-to-end SIMP-120 wall time versus number of elements for the
           cantilever benchmark (log--log).
            The plotted points are the representative single-run timings from
            the dedicated scaling workflow used for
            Table~\ref{tab:simp_scaling}; Section~\ref{sec:results:stats}
            reports the separate five-repeat variability study at 216\,k and
            512\,k.
            All three paths show near-linear scaling ($\Ocal(\nelem)$).
            The fused FP32 path achieves 4.6--7.3$\times$ speedup over FP64;
            the larger gap at 2\,M and 4.9\,M is an observed trend for which the
            paper does not report cache-counter evidence.}
  \label{fig:simp_scaling}
\end{figure}

\begin{figure}[!htbp]
  \centering
  \includegraphics[width=0.85\linewidth]{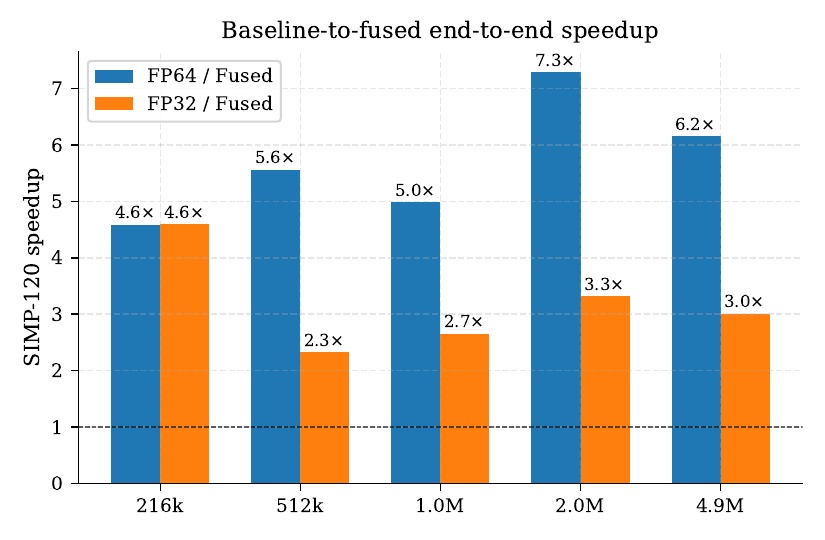}
\caption{Baseline-to-fused SIMP-120 speedup at five element counts.
           The blue bars report FP64/Fused and the orange bars report FP32/Fused.
           The bars are computed from the representative single-run scaling rows
           in Table~\ref{tab:simp_scaling}; Section~\ref{sec:results:stats}
           gives the separate repeat-study variability at 216\,k and 512\,k.
           FP64/Fused grows from $4.6\times$ at 216\,k to $7.3\times$
           at 2\,M and $6.2\times$ at 4.9\,M.}
  \label{fig:speedup_bars}
\end{figure}

\begin{figure}[!htbp]
  \centering
  \includegraphics[width=0.85\linewidth]{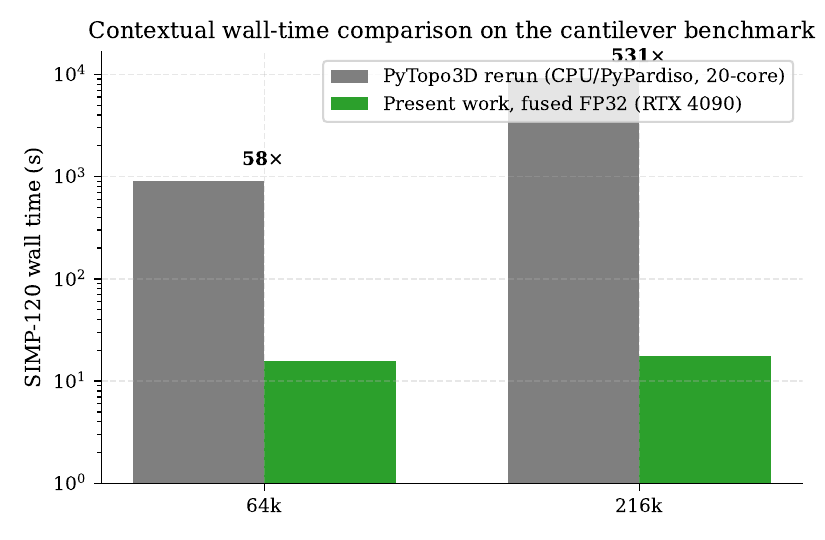}
  \caption{Wall-time comparison against local CPU/PyPardiso PyTopo3D
           reruns~\cite{pytopo3d2025} on the cantilever geometry and mesh sizes.
           This is a timing-only comparison: the two implementations use
           different SIMP formulations and mesh-scaling conventions, so
           compliance values are not directly comparable.
           The deposited PyTopo3D rerun log records 20 CPU cores and
           69.5\,GB available RAM, and the accompanying environment manifest
           records PyTopo3D 0.1.0, PyPardiso 0.4.7, and default host scheduling
           with no explicit thread pinning, so the figure is contextual only.
           The 64\,k GPU bar comes from a separate deposited 64\,k fused rerun;
           the 216\,k GPU bar comes from the main cantilever scaling ladder.
           The 64\,k and 216\,k bars use completed 120-iteration runs; the
           512\,k PyTopo3D run was terminated before completion and is excluded
           from this bar chart.}
  \label{fig:external_bars}
\end{figure}

\noindent\textbf{External CPU baseline comparison.}
To provide limited timing context against a publicly available Python-based
3D SIMP code, \Cref{fig:external_bars} compares the fused RTX\,4090 path with
local PyTopo3D reruns under its CPU/PyPardiso configuration~\cite{pytopo3d2025}
on the same cantilever geometry and mesh sizes.
This is a timing-only comparison: the two implementations use different SIMP
formulations and mesh-scaling conventions, so compliance values are not directly
comparable.
The deposited PyTopo3D rerun log records 20 CPU cores and 69.5\,GB available
RAM for that host, and the accompanying environment manifest records
PyTopo3D 0.1.0, PyPardiso 0.4.7, and default host scheduling with no explicit
thread pinning; the figure should therefore be read as contextual timing
evidence rather than as a version-controlled head-to-head benchmark.
At 64\,k and 216\,k elements, the reported PyTopo3D rows use completed
120-iteration runs; the fused GPU path is faster by approximately $58\times$ and
$531\times$, respectively, but these ratios should be read only as local timing
context rather than as a formulation-matched benchmark.

\begin{figure}[!htbp]
  \centering
  \includegraphics[width=\linewidth]{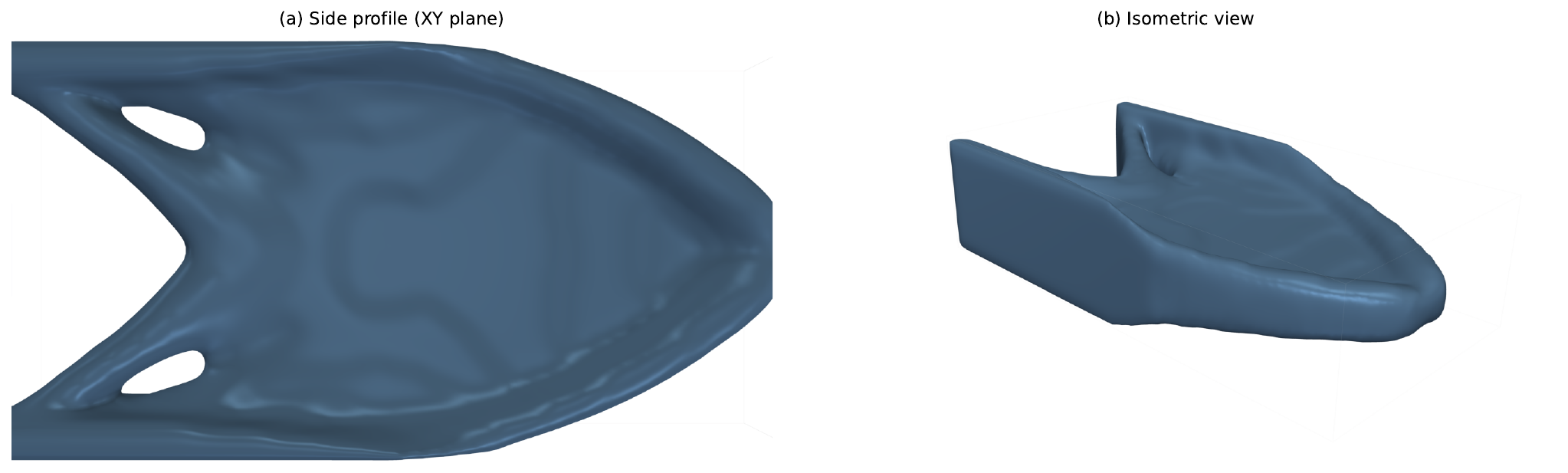}
\caption{Selected cantilever topology, $120\times 60\times 30 = 216{,}000$
            elements (selected design iterate from a separate same-configuration fused FP32
            SIMP-120 rerun,
            $V_f=0.30$; smoothed marching-cubes isosurface at $\rho=0.5$
            with 80-step Laplacian smoothing, relaxation factor 0.08).
           \emph{Left}: side-profile view (XY plane) showing the classic
           arrowhead-arch truss with two oval cutouts.
           \emph{Right}: isometric view revealing the three-dimensional
           ribbed shell structure.
           Grayness~$= 0.000$; compliance~$= 2.174$.}
  \label{fig:cantilever_render}
\end{figure}

\begin{figure}[!htbp]
  \centering
  \includegraphics[width=\linewidth]{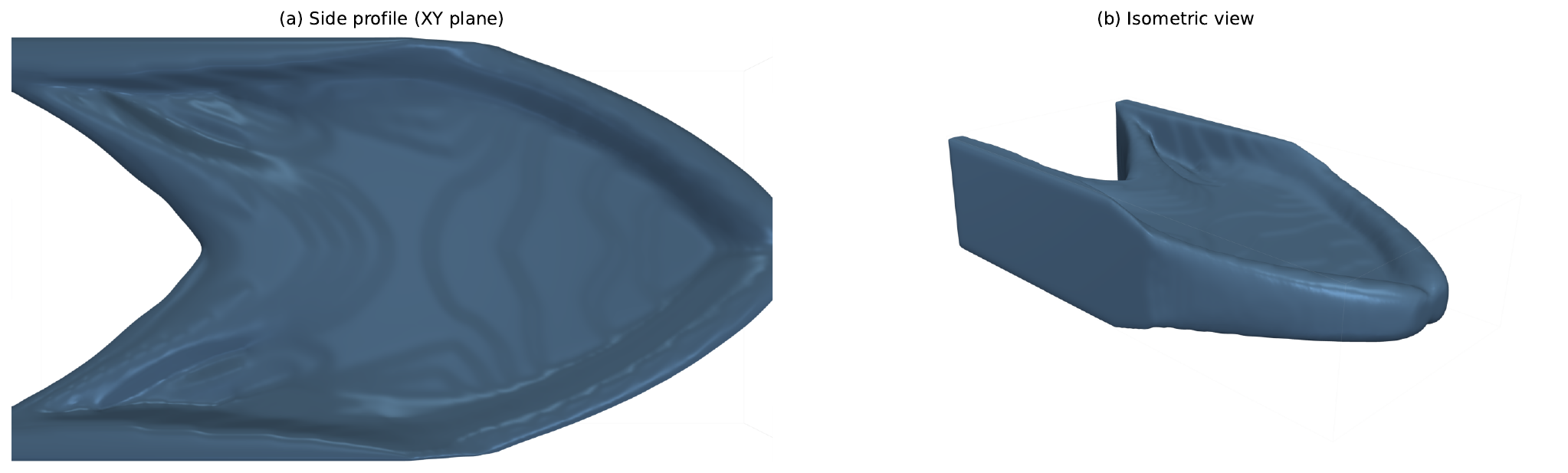}
\caption{Selected cantilever topology at $200\times 100\times 50 = 1{,}000{,}000$
            elements (selected design iterate from a separate same-configuration fused FP32
            SIMP-120 rerun,
            $V_f=0.30$; smoothed marching-cubes isosurface at $\rho=0.5$
            with 80-step Laplacian smoothing, relaxation factor 0.08).
           \emph{Left}: side-profile view. \emph{Right}: isometric view.
            The higher resolution resolves a cleaner arrowhead arch at the same
            qualitative design family; the corresponding table-supporting fused
            run in \Cref{tab:simp_scaling} takes 45.4\,s.
            Grayness~$= 0.000$; compliance~$= 1.608$.}
  \label{fig:cantilever_1m_render}
\end{figure}

\FloatBarrier
\subsection{Torsion Benchmark}
\label{sec:results:torsion}

The torsion benchmark serves as a difficult stress test with much higher CG iteration
counts than the cantilever.
\Cref{tab:torsion} summarizes results for the $165\times 55\times 55 = 499{,}125$
element mesh.

\begin{table}[h]
\centering
\caption{SIMP-120 torsion benchmark ($\nelem = 499{,}125$, RTX\,4090).
         FP64 and FP32 denote the three-stage baselines.
         Reported compliance is the selected compliance from the
         120-step continuation schedule.
         In the reported histories, 108--110 of the 120 iterations hit the
         current 1{,}000-iteration CG cap.}
\label{tab:torsion}
\begin{tabular}{lrrrr}
\toprule
Path & Wall time (s) & VRAM (GB) & Compliance & Speedup vs.\ FP64 \\
\midrule
FP64   & 643.5 & 2.46 & 2.0357 & 1.0$\times$ \\
FP32   & 415.0 & 2.32 & 2.0360 & 1.6$\times$ \\
Fused  & 146.5 & 2.38 & 2.0360 & \textbf{4.4$\times$} \\
\bottomrule
\end{tabular}
\end{table}

The torsion problem requires substantially more CG iterations per SIMP step than
the cantilever at a comparable size, which is why the absolute wall times are
larger than for the cantilever at a comparable 512\,k scale.
In fact, 108 of 120 FP64 iterations and 110 of 120 FP32/fused iterations reach the
current 1{,}000-iteration cap, with the first cap appearing at iteration 2 in all
three histories.
The torsion history should therefore be read as a difficult capped stress test
rather than as a clean low-iteration regime.
Even with that harder linear-solve profile, the fused kernel still delivers a
4.4$\times$ end-to-end speedup over the FP64 torsion baseline.

\begin{figure}[!htbp]
  \centering
  \includegraphics[width=0.72\linewidth]{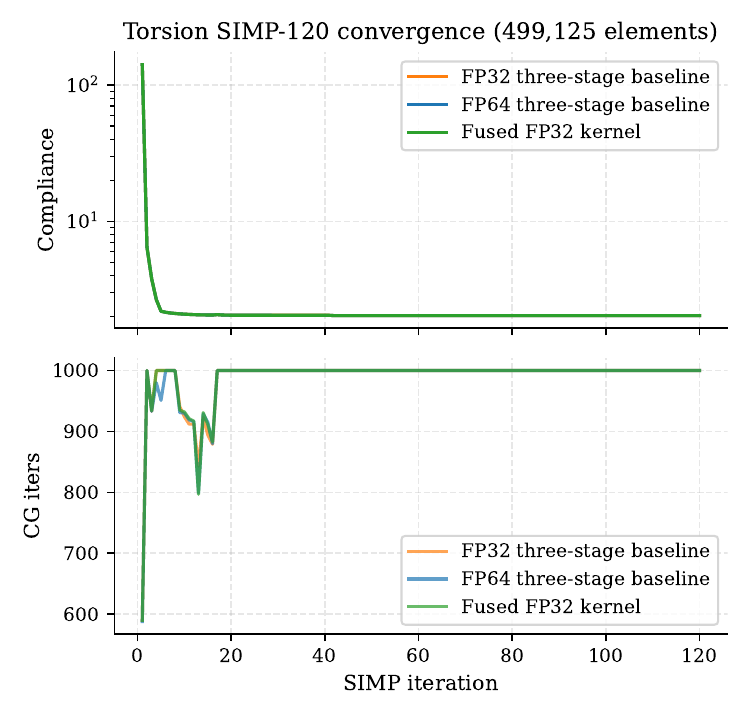}
  \caption{SIMP-120 convergence history for the 499{,}125-element torsion benchmark.
            \emph{Top}: compliance versus SIMP iteration (log scale) --- all three
            paths follow nearly identical trajectories throughout the continuation.
            \emph{Bottom}: CG iteration count per SIMP step --- the high CG
            counts reflect the elevated difficulty of the torsion linear systems,
            and 108--110 of the reported iterations hit the current 1{,}000-iteration cap.
            The compliance axis uses a log scale to show the full trajectory over
            the large early-to-late drop.}
  \label{fig:torsion_convergence}
\end{figure}

\begin{figure}[!htbp]
  \centering
\includegraphics[width=0.84\linewidth]{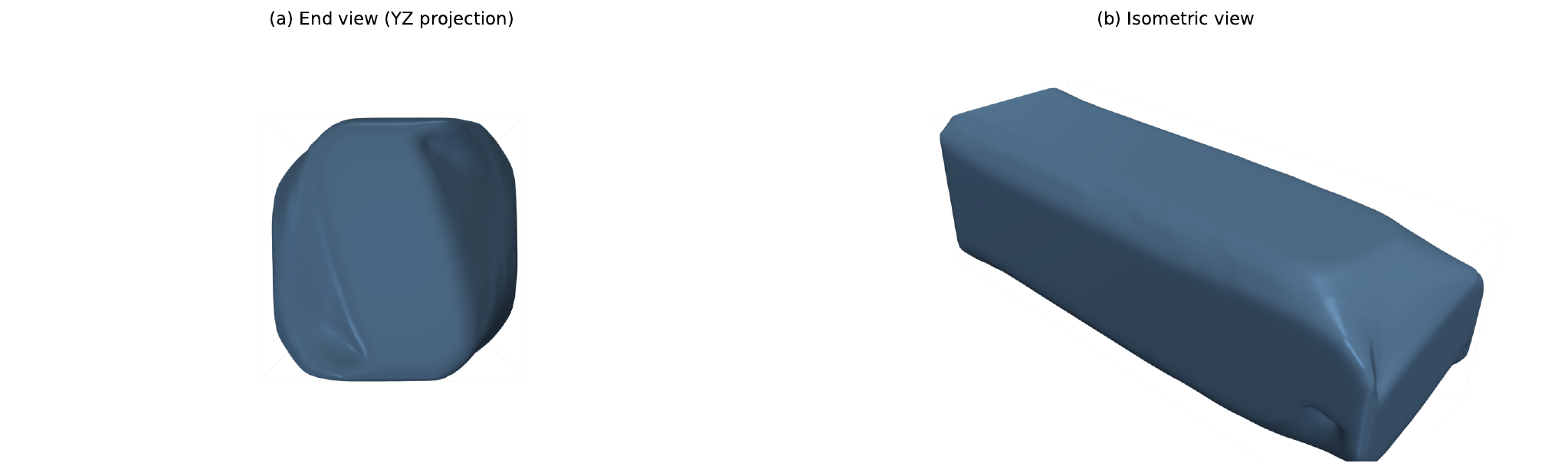}
\caption{Selected topology for the 499{,}125-element torsion shaft benchmark
            (selected design from a separate same-configuration fused FP32
            SIMP-120 rerun,
            $\vf = 0.25$; smoothed marching-cubes isosurface at $\rho=0.5$
            with 80-step Laplacian smoothing, relaxation factor 0.08).
            \emph{Left}: end-view projection (YZ view) of the selected density field.
            \emph{Right}: isometric surface rendering of the same selected field.
            This smoothed isosurface is a qualitative shape illustration rather than
            direct evidence of the internal cavity geometry.
            Grayness $= 4.1\times10^{-5}$;
            selected compliance $= 2.0360$.}
  \label{fig:torsion_render}
\end{figure}

\FloatBarrier
\subsection{MBB and Bridge Hard-Problem Stress Test}
\label{sec:results:bridge}

\begin{table}[H]
\centering
\caption{Hard-problem stress test on the RTX\,4090 ($\nelem = 187{,}500$).
         FP64 and FP32 denote the three-stage baselines.
         The FEA-only rows report mean wall time over five timed cold-start
         solves after warmup.
         The SIMP-60 cold-start rows report single representative 60-iteration
         runs without warm-start.
         For the FEA-only rows, compliance parity is computed from the final
         uniform-density cold-start solve compliance relative to FP64.
         For the SIMP-60 rows, it is computed from the selected best-valid
         compliance relative to FP64.}
\label{tab:bridge_hard}
\begin{tabular}{llrrrrr}
\toprule
Problem & Mode & Unit & FP64 & FP32 & Speedup & Compliance parity \\
\midrule
MBB & FEA-only & ms & 1138.0 & 754.7 & 1.51$\times$ & 2.61\% \\
Bridge & FEA-only & ms & 1130.8 & 890.0 & 1.27$\times$ & 0.49\% \\
MBB & SIMP-60 cold-start & s & 74.13 & 49.75 & 1.49$\times$ & 1.74\% \\
Bridge & SIMP-60 cold-start & s & 75.24 & 49.78 & 1.51$\times$ & 0.16\% \\
\bottomrule
\end{tabular}
\end{table}

The hard-problem study includes both the bending-dominated MBB case and the
distributed-load bridge case at $150\times 50\times 25 =
187{,}500$ elements.
These two presets share the same mesh size and Jacobi-PCG implementation, but
they are not like-for-like boundary-condition matches: MBB uses $\vf=0.50$
with a concentrated point load and mixed pin constraints, whereas the bridge
uses $\vf=0.30$ with fixed/roller supports and a distributed top load.
The rows should therefore be read as two separate hard-case probes rather than
as a controlled MBB-versus-bridge comparison.
In the FEA-only rows, all four cold-start solves hit the current
1{,}000-iteration cap; the SIMP-60 rows capture the short continuation runs.

The bridge-side rows indicate that the FP32-versus-FP64 three-stage trend is
not specific to the cantilever boundary condition in the single distributed-load
case tested here: the FP32 three-stage path remains faster than FP64 on both
the cold-start FEA solve and the truncated SIMP continuation, while the
compliance drift remains small. This evidence is limited to the three-stage
paths; the paper does not report a direct fused-kernel bridge benchmark.

\FloatBarrier
\subsection{Additional Qualitative Structures}
\label{sec:results:qualitative}

\begin{figure}[!htbp]
  \centering
  \includegraphics[width=\linewidth]{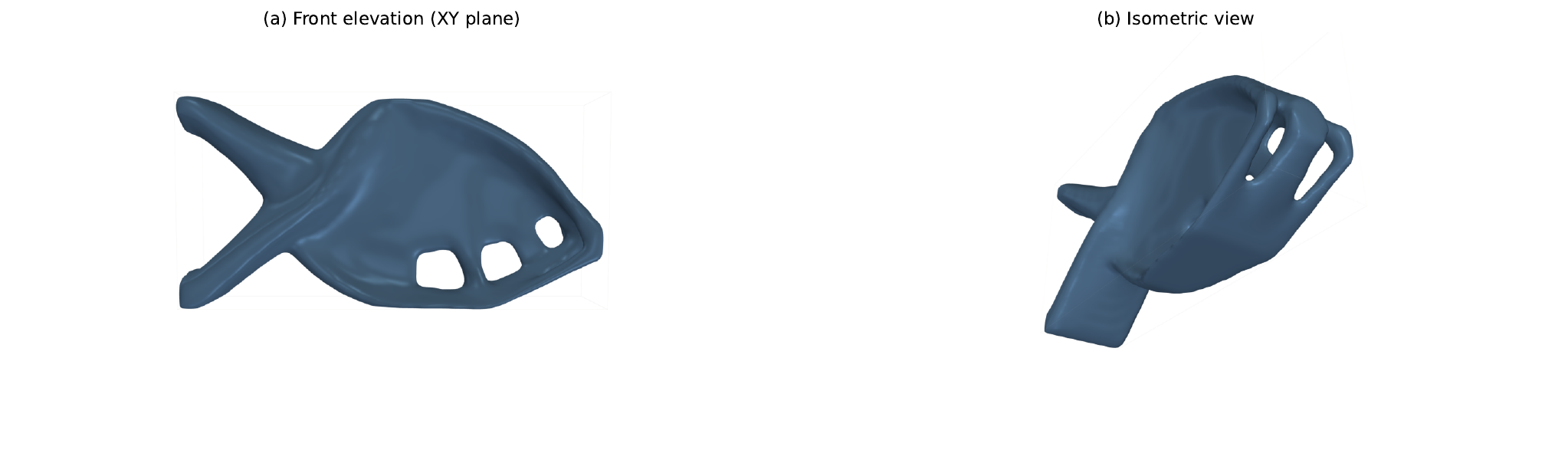}
  \caption{Separate qualitative exemplar from the bridge-family load class:
            selected topology for a left-clamped, right-roller beam with central load,
            $120\times 60\times 30 = 216{,}000$ elements (selected
            iterate from a separate fused FP32 SIMP-120 run with $V_f=0.30$ and
            initial filter radius $\rmin=3.0$ for this qualitative exemplar only;
            smoothed marching-cubes isosurface at $\rho=0.5$ with 80-step
            Laplacian smoothing, relaxation factor 0.08).
           The solver recovers an arch-and-strut system transferring the
           center-top load to the fixed left wall and the roller right support
           --- a structurally distinct topology from the quantitative cantilever,
           torsion, and 187{,}500-element bridge timing benchmarks, and a qualitative companion to the bridge
           hard-problem stress test reported above.
           This panel is qualitative only and is not the quantitative source for
           Table~\ref{tab:bridge_hard}.
           Grayness~$= 0.000$; compliance~$= 0.903$.}
  \label{fig:bridge_render}
\end{figure}

Figure~\ref{fig:bridge_render} shows a left-clamped, right-roller beam with
central load as a qualitative companion to the bridge stress test above.
This run is included to illustrate topology diversity; it is not the source of
\Cref{tab:bridge_hard} and should not be read as a duplicate quantitative
measurement.

An additional centered-patch cantilever topology at $\vf=0.10$ from a separate
fused-FP32 SIMP rerun is included in Appendix~\ref{app:repro} as a supplemental
qualitative example outside the main timing tables.

\noindent\textbf{Grayness metric.}
For the qualitative figures, grayness is reported using the definition given in
Section~\ref{sec:results:setup}; $g=0$ denotes a fully binary design and larger
values indicate more intermediate-density material.

\FloatBarrier
\subsection{BF16 Convergence Study}
\label{sec:results:bf16}

We investigate the convergence behavior of the BF16 WMMA kernel when integrated
into the CG solver via the reported FP32/BF16 residual-correction experiments.
\Cref{tab:bf16conv} reports compliance and relative error for four solver configurations
for a \emph{single cold-start linear solve} on the cantilever benchmark at 64\,k
and 216\,k elements, with uniform density $\rho=0.5$ and $\penal=3$.
These are not full SIMP runs; they are representative linear-solve
experiments designed to expose solver convergence behavior.

\begin{table}[h]
\centering
\caption{BF16 linear-solve convergence study for a single cold-start linear solve on the
         cantilever benchmark ($\rho=0.5$, $\penal=3$).
          FP32 is the reference.
          Plain BF16 CG uses no refinement.
          BF16-IR (inner tol $10^{-3}$) and BF16-IR (inner tol $10^{-5}$)
          denote BF16 inner solves with iterative-refinement outer loops
          (maximum 8 outer corrections) at two inner tolerances.
          Compliance relative error
          $\delta_c = |c_{\mathrm{ref}} - c| / c_{\mathrm{ref}}$.}
\label{tab:bf16conv}
\setlength{\tabcolsep}{5pt}
\begin{tabular}{llrrrr}
\toprule
Size & Solver & Wall (ms) & CG iters & Compliance & $\delta_c$ \\
\midrule
\multirow{4}{*}{64\,k}
  & FP32 reference      &  456 & 345 & 13.431 & --- \\
  & Plain BF16 CG       &  186 & 230 &  6.178 & 0.540 \\
  & BF16-IR (inner tol $10^{-3}$) & 1201 & 1085 &  7.014 & 0.478 \\
  & BF16-IR (inner tol $10^{-5}$) & 2144 & 1857 &  7.014 & 0.478 \\
\midrule
\multirow{4}{*}{216\,k}
  & FP32 reference      &  591 & 512 & 11.022 & --- \\
  & Plain BF16 CG       &  281 & 234 &  6.102 & 0.446 \\
  & BF16-IR ($10^{-3}$) & 1581 & 1257 &  6.424 & 0.417 \\
  & BF16-IR ($10^{-5}$) & 2646 & 2044 &  6.424 & 0.417 \\
\bottomrule
\end{tabular}
\end{table}

The results are consistent with the convergence analysis of
Section~\ref{sec:method:bf16conv}
in three ways.

First, plain BF16 CG produces compliance values with 44--54\% error,
indicating that the CG solver does not converge to a valid FE solution.
The compliance field is therefore numerically invalid, and BF16 CG is unusable as a direct solver
for the TO linear systems encountered here.

Second, BF16 iterative refinement partially mitigates but does not
resolve the problem: compliance error drops from 54\% to 48\% at 64\,k and from
45\% to 42\% at 216\,k, but stagnates there regardless of how tight the inner
tolerance is set.
Tightening the inner tolerance from $10^{-3}$ to $10^{-5}$ merely increases CG
iteration count (1,085 $\to$ 1,857 at 64\,k) and wall time (about $1.7$--$1.8\times$)
without
improving the outer residual---a clear signature of the stagnation predicted by
the $\eps_{\mathrm{BF16}}\cdot\kappa(\Kmat) \geq 1$ barrier.

Third, the stagnation level is problem-size-invariant: at both 64\,k and 216\,k,
BF16-IR converges to the same compliance plateau ($\delta_c \approx 0.42$--$0.48$),
indicating that the dominant conditioning effect is more geometry- and
penalization-driven than mesh-size-driven.
Because the test state is the uniform-density configuration $(\rho=0.5,\penal=3)$,
it represents a lower-contrast probe rather than the hardest late-continuation SIMP
system; the companion kappa estimation study in Section~\ref{sec:results:kappa}
shows the threshold crossing already at all tested uniform states and both
reported penalization levels, directly supporting this interpretation.

In the single-solve cold-start regime tested here, these findings indicate that
the BF16 WMMA kernel is not a viable drop-in replacement for FP32/FP64 in the
CG solve for the tested 3D SIMP systems, and provide a focused empirical
characterization of this failure mode.

\FloatBarrier
\subsection{Condition Number Estimation}
\label{sec:results:kappa}

To ground the BF16 convergence analysis in a direct measurement rather than an
inference from stagnation behavior, the paper includes
a dedicated condition-number estimation workflow, which estimates
$\kappa(\Kmat)$ via matrix-free power iteration~\cite{golub2013matrix}
(50 fixed steps for the largest eigenvalue in the current study) and
inverse iteration (5--6 steps for the smallest eigenvalue).
Both extremal eigenvalues are estimated entirely through matrix-free
$\Kmat\mathbf{v}$ applications (Algorithm~\ref{alg:fused}), so no explicit
$\Kmat$ is assembled and the procedure is feasible at all tested mesh sizes.
The current workflow uses fixed start vectors (seed 42 for power iteration and
seed 123 for inverse iteration), and each inverse-iteration inner solve uses
SciPy CG with relative and absolute tolerance $10^{-8}$ and a
2{,}000-iteration cap.
The reported tabulation covers the
uniform-density initialization ($\rho=0.5$) at two penalization levels
($\penal \in \{3.0, 5.0\}$) for 64\,k, 216\,k, and 512\,k elements.

The reported estimates already indicate that $\kappa(\Kmat)$ far exceeds the BF16
stability threshold at the lowest-contrast test state (uniform density,
$\rho=0.5$).
At 64\,k elements, power iteration yields
$\kappa \approx 6.1\times10^{5}$, giving
$\varepsilon_{\mathrm{BF16}}\cdot\kappa \approx 2.4\times10^{3}$---nearly
$9\times$ above the $1/\varepsilon_{\mathrm{BF16}} = 256$ convergence threshold.
At 216\,k elements the conditioning is larger:
$\kappa \approx 1.3\times10^{6}$ and
$\varepsilon_{\mathrm{BF16}}\cdot\kappa \approx 5.2\times10^{3}$---more than $20\times$
above the threshold.
At 512\,k elements the trend continues:
$\kappa \approx 2.3\times10^{6}$ and
$\varepsilon_{\mathrm{BF16}}\cdot\kappa \approx 9.1\times10^{3}$---more than $35\times$
above the threshold.
Because all six reported rows hit the 50-step power-iteration cap, these
$\kappa$ values should be read as conservative estimates for the corresponding
uniform states.
The three estimates scale approximately as
$\mathcal{O}(n_{\mathrm{elem}}^{2/3})$ ($\kappa$ ratios of $2.18\times$ for
the 216\,k/64\,k pair and $1.75\times$ for the 512\,k/216\,k pair, consistent
with the theoretical $h^{-2}$ growth for elliptic PDEs on uniform meshes),
indicating that the conditioning barrier intensifies monotonically with mesh
refinement.
All three estimates are insensitive to the penalization exponent at uniform
density (identical $\kappa$ for $\penal=3$ and $\penal=5$): at $\rho_e=0.5$
all element moduli are equal regardless of $\penal$, so the condition number
is driven solely by the mesh geometry.
At late-SIMP density contrasts, near-void elements
($\rho_e \approx \rho_{\min} = 10^{-9}$) introduce columns with modulus ratio
$\rho_{\min}^{\penal} \approx 10^{-27}$, driving $\kappa$ orders of magnitude
higher still.
These uniform-density measurements already exceed the BF16 stability threshold by
wide margins. Late-SIMP states are expected to be more ill-conditioned still,
but that stronger statement is an inference rather than a direct measurement
here. A separate uniform-density BF16 extension at $\penal = 4.5$ shows the same
qualitative stall as at $\penal = 3.0$ ($\delta_c \approx 0.54$, BF16;
$\delta_c \approx 0.48$, BF16-IR), with no evidence of improvement at the
higher penalization.

\subsection{Statistical Reproducibility and Run-to-Run Determinism}
\label{sec:results:stats}

The end-to-end SIMP tables in the main text report single representative runs.
To quantify run-to-run variability, a dedicated repeat-run workflow
runs SIMP-120 $N=5$ times at each (size, path) pair for the cantilever benchmark
and reports mean wall time, standard deviation, and coefficient of variation
(CV\%).
Table~\ref{tab:repeat_study} reports those deposited repeat-run summaries.

\begin{table}[H]
\centering
\caption{Five-repeat SIMP-120 variability study for the cantilever benchmark.
         Each row summarizes the deposited $N=5$ reruns at a fixed
         (size, path) pair.}
\label{tab:repeat_study}
\begin{tabular}{llrrrrr}
\toprule
Size & Path & $N$ & \makecell{Wall mean\\$\pm$ std (s)} & \makecell{CV\\(\%)} & $\bar{c}$ & $c_{\mathrm{std}}$ \\
\midrule
216\,k & FP64  & 5 & $66.39 \pm 6.30$   & 9.49  & 2.172194 & $3.42\times10^{-6}$ \\
216\,k & Fused & 5 & $39.58 \pm 11.82$  & 29.85 & 2.173862 & $1.26\times10^{-5}$ \\
512\,k & FP64  & 5 & $111.45 \pm 9.34$  & 8.38  & 1.825360 & $9.10\times10^{-6}$ \\
512\,k & Fused & 5 & $26.48 \pm 0.48$   & 1.82  & 1.827701 & $2.76\times10^{-5}$ \\
\bottomrule
\end{tabular}
\end{table}

Table~\ref{tab:repeat_study} shows near-zero compliance
variation ($c_\mathrm{std} \leq 3\times10^{-5}$, less than $0.002$\% of the mean
compliance) across all four tested (size, path) combinations,
indicating negligible run-to-run compliance variation in the tested cases.
Wall-time CV is larger: the FP64 baseline shows 9.5\% at 216\,k
($66.4\pm6.3$\,s) and 8.4\% at 512\,k ($111.5\pm9.3$\,s), reflecting
OS-level scheduling and thermal jitter between cold-start reinitializations.
The fused path shows only 1.8\% CV at 512\,k ($26.5\pm0.5$\,s);
the fused kernel was already compiled by the preceding 216\,k run within the same
process, so all five 512\,k reps represent pure post-JIT execution.
At 216\,k the fused CV rises to 29.9\% ($39.6\pm11.8$\,s): the first cold-start
rep includes CuPy NVRTC kernel-compilation overhead, but the observed five-run
spread also contains later slow reps, so this row should be read as a mixed
cold-start/thermal-variability measurement rather than a pure post-JIT timing
study.
The representative single-run values in Table~\ref{tab:simp_scaling} come from a
separate scaling workflow rather than from this repeat study. In that workflow,
each size is executed sequentially as FP64, FP32, then fused within one process,
so Table~\ref{tab:simp_scaling} should be read as a representative scaling
trajectory, whereas the present table quantifies run-to-run variability.

A companion determinism workflow runs
$N=10$ identical cold-start fused FP32 solves on the same problem and measures
the relative compliance spread
$\Delta c / c_{\mathrm{ref}} = (c_{\max} - c_{\min}) / c_{\mathrm{FP64}}$.
This study quantifies non-determinism from the FP32 atomic
scatter in Algorithm~\ref{alg:fused}, whose floating-point reordering across
GPU threads is theoretically non-associative~\cite{higham2002accuracy}.
Table~\ref{tab:determinism} reports the deposited determinism summaries.

\begin{table}[H]
\centering
\caption{Determinism study for repeated cold-start fused FP32 solves on the
         cantilever benchmark from the dedicated determinism workflow (separate
         from the representative residual-history solve shown in
         Figure~\ref{fig:cg_residual}). Here
         $\Delta c / c_{\mathrm{FP64}} = (c_{\max} - c_{\min}) / c_{\mathrm{FP64}}$
         and $\delta_c = |c_{\mathrm{fused}} - c_{\mathrm{FP64}}| / c_{\mathrm{FP64}}$.}
\label{tab:determinism}
\begin{tabular}{lrrrrrr}
\toprule
Size & $N$ & $c_{\min}$ & $c_{\max}$ & $\Delta c / c_{\mathrm{FP64}}$ & $\max \delta_c$ & CG iters \\
\midrule
64\,k  & 10 & 13.431176 & 13.431200 & $1.78\times10^{-6}$ & $5.94\times10^{-4}$ & 345 \\
216\,k & 10 & 11.022279 & 11.022296 & $1.56\times10^{-6}$ & $1.09\times10^{-3}$ & 512--513 \\
\bottomrule
\end{tabular}
\end{table}

Table~\ref{tab:determinism} shows
run-to-run compliance spread of $1.78\times10^{-6}$ relative at 64\,k elements
and $1.56\times10^{-6}$ at 216\,k elements---both well below $10^{-4}$---indicating
that the fused FP32 path is deterministic to well within engineering accuracy
in practice despite the theoretically non-associative atomics.
The FP32-versus-FP64 absolute compliance offset ($\delta_c \approx 6\times10^{-4}$
at 64\,k and $1.1\times10^{-3}$ at 216\,k) arises from the precision difference
itself and is consistent with the values reported in Table~\ref{tab:simp_scaling}.

\begin{sloppypar}
\noindent\textbf{Selected-iterate validity.}
A further validation study
compares each path's selected compliance from the standard 1{,}000-iteration
CG cap against a high-cap 5{,}000-iteration rerun.
\end{sloppypar}
This validation currently covers only the 216\,k and 512\,k FP64 and fused
cantilever rows.
Table~\ref{tab:highcap_validation} reports the deposited high-cap reruns.

\begin{table}[t]
\centering
\caption{Selected-iterate high-cap validation for the cantilever benchmark.
         Canonical rows use the current 1{,}000-iteration CG cap; the validation
         reruns raise that cap to 5{,}000. Relative difference is
         $|c_{\mathrm{canonical}} - c_{\mathrm{high}}| / c_{\mathrm{canonical}}$.}
\label{tab:highcap_validation}
\begin{tabular}{llrrrrr}
\toprule
Size & Path & Canonical cap & High cap & $c_{\mathrm{canonical}}$ & $c_{\mathrm{high}}$ & Relative difference \\
\midrule
216\,k & FP64  & 1000 & 5000 & 2.172196 & 2.172196 & $3.61\times10^{-9}$ \\
216\,k & Fused & 1000 & 5000 & 2.173877 & 2.173838 & $1.78\times10^{-5}$ \\
512\,k & FP64  & 1000 & 5000 & 1.825346 & 1.824751 & $3.26\times10^{-4}$ \\
512\,k & Fused & 1000 & 5000 & 1.827643 & 1.827066 & $3.16\times10^{-4}$ \\
\bottomrule
\end{tabular}
\end{table}

Reported results show a maximum deviation
of $|\Delta c|/c \leq 3.26\times10^{-4}$ across all four (size, path)
combinations tested (216\,k and 512\,k; FP64 and Fused):
the 216\,k FP64 canonical and high-cap compliances are identical to
$3.6\times10^{-9}$ relative error; the worst case (512\,k FP64) is $3.26\times10^{-4}$, while
the 512\,k fused row is $3.16\times10^{-4}$.
All four cases fall below the current $5\times10^{-4}$ acceptance threshold,
indicating that the selected compliances reported in the tested 216\,k and
512\,k FP64/fused rows of Table~\ref{tab:simp_scaling} are not artifacts of the
1{,}000-iteration cap but represent well-converged optima for the tested SIMP
continuation schedule. The larger 2\,M and 4.9\,M scaling rows are not part of
this high-cap validation set.

\FloatBarrier
\subsection{VRAM Scaling and Memory Efficiency}
\label{sec:results:vram}

Table~\ref{tab:vram} summarizes end-of-run VRAM allocation snapshots at select
sizes.

\begin{table}[h]
\centering
\caption{End-of-run VRAM allocator snapshots on the RTX\,4090 (24\,GB total).
         Values are the deposited used-memory snapshots recorded immediately
         after the canonical SIMP scaling runs; they are not instrumented
         peak-memory measurements.
         The fused path is modestly more memory-efficient than the three-stage
         FP64 baseline, but the observed savings depend on allocator behavior
         and on other live buffers in the SIMP loop.}
\label{tab:vram}
\begin{tabular}{lrrrr}
\toprule
$\nelem$ & FP64 (GB) & FP32 (GB) & Fused (GB) & Fused saving \\
\midrule
   216\,k &  1.93 & 1.86 & 1.89 & $0.04$\,GB \\
  512\,k &  2.42 & 2.35 & 2.41 & $0.01$\,GB \\
  1\,M &  3.43 & 3.14 & 3.25 & $0.18$\,GB \\
  2\,M &  5.09 & 5.13 & 5.00 & $0.09$\,GB \\
 4.9\,M & 11.06 & 10.39 & 10.07 & $0.99$\,GB \\
\bottomrule
\end{tabular}
\end{table}

These snapshots grow approximately linearly with $\nelem$ for all paths, as
expected from the $\Ocal(\nelem)$ matrix-free memory model.
The fused path achieves a modest end-of-run saving versus the three-stage FP64
baseline by eliminating the two 24-component per-element work arrays; in the
deposited 4.9\,M run the observed difference is about $1$\,GB.
The total footprint is not dominated by any single small array:
persistent buffers include the global DOF state, the element-to-DOF table, the
density field, the sparse density-filter operators, and solver/preconditioner
workspace, with the exact split depending on path and allocator reuse.
The RTX\,4090's 24\,GB budget is consistent with the deposited 8\,M-element
FEA-only allocator snapshot (10.56\,GB used at end-of-run), but that point
should be read as a capped memory-allocation stress test rather than as a
peak-memory guarantee or a fully converged solve.

\FloatBarrier
\subsection{Correctness Verification}
\label{sec:results:correctness}

Numerical correctness of the fused FP32 kernel is assessed by three cross-checks.

\begin{figure}[!htbp]
  \centering
  \includegraphics[width=0.80\linewidth]{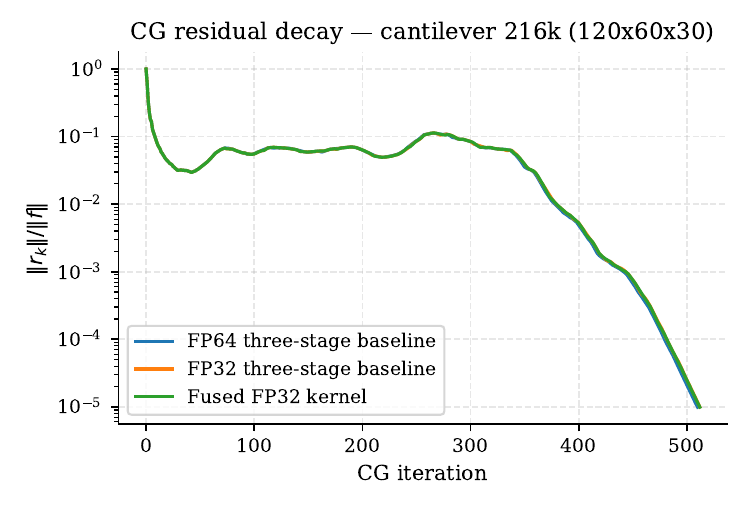}
  \caption{Representative CG residual decay for a cold-start linear solve at
           uniform density $\rho=0.5$ on the
           $120\times 60\times 30 = 216{,}000$-element cantilever benchmark.
           The plotted quantity is the unpreconditioned relative residual
           $\|\mathbf{r}_k\|/\|\mathbf{f}\|$.
           All three solver paths (FP64, FP32, fused FP32) follow nearly
           identical trajectories.}
  \label{fig:cg_residual}
\end{figure}

\begin{enumerate}[nosep]
  \item \textbf{Compliance parity.}
        Across the reported tables, the fused FP32 selected-summary compliance
        remains close to the FP64 reference, with sub-0.2\% deviations through
        1\,M elements, a 2.46\% deviation at 2\,M, and a 0.12\% deviation at
        4.9\,M. The high-cap selected-iterate validation currently covers only
        the 216\,k and 512\,k cantilever rows (Section~\ref{sec:results:stats}).
  \item \textbf{Selected-field grayness.}
        The reported render metadata for the selected fused density fields shown in
        Figures~\ref{fig:cantilever_render}, \ref{fig:cantilever_1m_render},
        and \ref{fig:torsion_render} indicate essentially discrete designs
        ($g \approx 0$), indicating that the fused solver reaches the same
        low-grayness regime as the reference paths.
  \item \textbf{Residual monitoring.}
        Figure~\ref{fig:cg_residual} shows a representative 216\,k cold-start
        linear solve at uniform density $\rho=0.5$.
        The FP64, FP32, and fused traces are nearly indistinguishable and terminate
        in 510--512 CG iterations, indicating that the fused operator introduces no
        visible additional error in this representative case. The supplementary
        trace file stores the exact per-iteration residual histories for this
        representative solve.
\end{enumerate}

\section{Discussion}
\label{sec:discussion}

\subsection{Interpretation of Speedup Trends}
\label{sec:discussion:speedup}

The fused gather--GEMM--scatter kernel achieves 6.0--6.6$\times$ per-matvec
speedup (synthetic hot-path microbenchmark) and end-to-end SIMP-120 wall-time
speedup of 4.6--7.3$\times$ on cantilever plus 4.4$\times$ on the
499{,}125-element torsion benchmark over the FP64 three-stage baseline.
Against the same-precision FP32 three-stage path, the fused solver delivers
2.3--4.6$\times$ on cantilever and 2.8$\times$ on torsion.
CUDA-event measurements on the actual cantilever operator
(Table~\ref{tab:bw_utilization}) yield isolated per-call speedups of
\textbf{8.9--13.8$\times$}---higher than the synthetic benchmark. That gap is
consistent with the extra Python/CuPy dispatch overhead avoided when three API
calls are collapsed into one launch; the synthetic benchmark reports
pure GPU kernel time and thus captures only the hardware-level benefit.
Both measurements indicate that the operator is memory-bandwidth bound:
the fused kernel achieves 61--179\,GB/s effective bandwidth (6--18\% of
the 1,008\,GB/s peak), versus 20--43\,GB/s (2--4\% of peak) for the
FP64 baseline---both well below the roofline ridge point of
$\approx$82\,FLOP/B.
The per-matvec speedup is consistent with two mechanisms that compound.
The primary mechanism is \emph{DRAM traffic elimination}: the fused kernel avoids
writing the intermediate per-element work arrays ($\mathbf{u}_{\mathrm{elem}}$ and
$\mathbf{f}_{\mathrm{elem}}$) to DRAM between stages, reducing the effective DRAM
load according to the profiling-guided traffic accounting in
Section~\ref{sec:results:profiling}.
The secondary mechanism is \emph{kernel launch overhead elimination}: the baseline
requires three separate GPU-resident stages (gather, batched GEMM,
reduction scatter), each incurring dispatch and synchronization overhead that
accumulates over hundreds of CG iterations per SIMP step.
At 64\,k elements, where the kernel execution time is short ($\sim$28--168\,$\mu$s
per stage), this overhead is proportionally large, and its elimination accounts for
a non-trivial fraction of the total speedup.

The higher relative speedup at 2\,M versus 1\,M elements (7.3$\times$ vs.\
5.0$\times$) may appear counterintuitive since both sizes share the same kernel
architecture.
One plausible explanation is cache effects: the $\Ke^{\mathrm{unit}}$ matrix
(2.25\,kB in the FP32 fused path; 2.0\,kB for the BF16 padded shared-memory
tile in the WMMA path) is broadcast from L2 cache
to all 2\,M thread blocks at runtime; at very large sizes the sheer number of active
thread blocks keeps the L2 warm, while at moderate sizes ($\sim$1\,M) there is a
transition region where the cache-miss penalty from loading $\Ke^{\mathrm{unit}}$
is visible but not yet amortized over enough active blocks.
The paper does not include cache-counter measurements, so this
interpretation should be read as a plausible explanation rather than as a
directly profiled mechanism, and we therefore treat it as an open performance
question rather than as a validated cache account.

The FP32 three-stage path achieves about 1.9--2.4$\times$ speedup over FP64 for large
sizes (512\,k--4.9\,M).
This is consistent with a bandwidth-bound regime in which FP32 halves the
per-entry data footprint relative to FP64, thereby reducing effective DRAM traffic,
rather than with the RTX\,4090's much larger FP32:FP64 peak-FLOP ratio.
At 216\,k, the FP32 path shows essentially no difference relative to FP64
(80.5\,s vs.\ 80.4\,s)
because the non-FEA SIMP overhead is still large enough to mask that matvec-level
FP32 advantage in the full end-to-end timing.

The bridge hard-problem stress test shows the same FP32-versus-FP64 three-stage
trend under one distributed-load case: 1.27$\times$ speedup for the cold-start FEA solve and
1.51$\times$ for the 60-iteration SIMP run, with compliance parity staying
within 0.5\% and 0.2\%, respectively.
That result does not directly benchmark the fused kernel on the bridge case, but
it is consistent with the same FP32-versus-FP64 three-stage trend on the tested
bridge case; direct fused-kernel bridge benchmarking remains future work.

\subsection{BF16 Tensor Cores: Promise and Precision Barrier}
\label{sec:discussion:bf16}

The reported profiling study includes a separate $14.3\times$
BF16-versus-FP64 GEMM proxy timing at 512\,k elements (80.6\,$\mu$s vs.\ 1154.3\,$\mu$s for FP64
batched DGEMM), yet
the full-pipeline speedup saturates near 6.7$\times$---nearly the same as the fused
FP32 kernel.
This result reveals an important architectural reality of the gather--GEMM--scatter
pattern: the gather and scatter stages, which involve globally irregular memory
accesses (driven by the DOF index table) and atomic reductions, are
\emph{not accelerated by tensor cores}.
These stages impose an irreducible memory-bandwidth cost that sets the throughput
ceiling regardless of how fast the GEMM stage runs.
In the current fused kernel architecture, gather + scatter account for
$\sim$196--1775\,$\mu$s of the FP64 full-pipeline time across the reported
64\,k--1\,M profiling rows, while
the GEMM stage accounts for 141--2281\,$\mu$s.
BF16 WMMA accelerates only the GEMM portion, so the total pipeline speedup is
bounded by the non-GEMM share. Using the reported 512\,k accounting argument from
Section~\ref{sec:results:profiling}, even a theoretically zero-cost GEMM would
still be capped near $8.6\times$ for that fused-BF16 row.

The BF16 convergence failure is a more fundamental concern.
Direct power-iteration estimates reported in
Section~\ref{sec:results:kappa} measure
$\kappa(\Kmat) \approx 6.1\times10^{5}$ at 64\,k,
$\approx 1.3\times10^{6}$ at 216\,k, and $\approx 2.3\times10^{6}$ at 512\,k,
for the uniform-density test state ($\rho=0.5$), giving
$\varepsilon_{\mathrm{BF16}}\cdot\kappa \approx 2.4\times10^{3}$, $5.2\times10^{3}$, and $9.1\times10^{3}$
respectively---well above the $1/\varepsilon_{\mathrm{BF16}} = 256$ sufficient threshold
implied by the standard IR bound $\varepsilon_{\mathrm{BF16}}\kappa(\Kmat) < 1$ from Carson and Higham~\cite{carson2018threeprecision}.
This places the tested systems firmly and deeply in the
$\varepsilon_{\mathrm{BF16}}\cdot\kappa(\Kmat) \gg 1$ regime where the IR guarantee does not apply.
Iterative refinement (IR) does not resolve this because IR's outer residual
correction step itself requires a matrix--vector product with $\Kmat$, and when
$\varepsilon_{\mathrm{BF16}}\cdot\kappa(\Kmat) \gg 1$, each inner BF16 solve introduces an error that
cannot be fully corrected by the outer FP32 residual-correction step.
In that regime, the BF16 solve is simply too inaccurate to reduce the outer
residual below the BF16 noise floor for the tested systems.

This result qualifies the optimistic projection of Henry et al.~\cite{henry2019bf16}
for the specific tested TO systems.
Henry et al.\ argue that BF16-IR converges ``over a large range of condition
numbers'' when FP32 is used for the outer refinement; our results show that this
claim does not extend to the reported 3D structural-elasticity experiments once the
effective condition number rises above the BF16 IR threshold.
The qualification is not a criticism of BF16-IR in general---for well-conditioned or
properly preconditioned systems (e.g., the lattice QCD operator studied by
Clark et al.~\cite{clark2010quda}, where mixed-precision Krylov solvers with
reliable updates succeed in practice)---but
it identifies the tested SIMP TO stiffness matrices as a regime where direct BF16-IR fails.

\subsection{Contextual Comparison to Prior Implementations}
\label{sec:discussion:comparison}

The closest published competitor is Tr{\"a}ff et al.~\cite{traff2023simple},
whose Futhark/OpenMP-C matrix-free solver achieves 65.5\,M elements on an NVIDIA
A100 (80\,GB HBM2e) with SSOR V-cycle multigrid preconditioning.
Our solver achieves 2\,M elements on an RTX\,4090 (24\,GB GDDR6X) with Jacobi
preconditioning with a few hundred CG iterations per cantilever SIMP step.
The two implementations are not directly comparable for three reasons.

First, the hardware differs: the A100 has 80\,GB HBM2e at up to 1.94\,TB/s versus
the 4090's 24\,GB GDDR6X at about 1\,TB/s \cite{nvidiaa1002025,nvidiaada2023}.
The VRAM gap contributes materially to the size difference, but it does not by
itself explain the full problem-size gap; the 4090's 24\,GB limits the reported
single-GPU study to $\sim$8\,M elements in the current FEA-only stress test.

Second, the preconditioner gap is responsible for the per-step efficiency
difference.
Tr{\"a}ff et al.\ report much lower per-step times on comparable A100-scale problems;
our Jacobi-PCG requires hundreds of iterations (per-step time 0.15--8\,s depending on size),
and the torsion stress test reaches the current 1{,}000-iteration limit in
108--110 of its 120 SIMP steps.
This paper should therefore be read as an implementation-focused single-GPU study,
not as a preconditioner-matched head-to-head benchmark.

Third, accessibility differs: Tr{\"a}ff et al.'s solver requires the Futhark
compiler or an additional build toolchain; our solver is Python-native once
CuPy is installed.
This accessibility advantage matters for the broad engineering community of TO
practitioners who work in Python ecosystems.

Wang et al.~\cite{wang2025matrixfree} achieve 128\,M elements on a 64\,GB CPU
workstation using geometric multigrid with non-dyadic Galerkin coarsening.
That result underscores the same point as Tr{\"a}ff et al.'s work: the main remaining
gap in the present implementation is preconditioning rather than matrix-free operator
throughput.

\subsection{Limitations}
\label{sec:discussion:limitations}

The present work has four principal limitations.

\noindent\textbf{Jacobi preconditioning.}
The Jacobi preconditioner limits CG convergence to $\Ocal(\sqrt{\kappa})$
iterations, which dominates wall time for problems with high condition numbers.
For the 3D MBB beam problem with standard pin-roller boundary conditions, the
near-rigid-body mode leads to very poor conditioning, and the accompanying
hard-problem MBB rows remain cap-limited under the implemented 1,000-iteration
Jacobi-PCG setting.
This means the current solver does not yet handle all standard TO benchmarks.
Geometric multigrid preconditioning would substantially reduce the per-step CG count and
is the most impactful single improvement available; it is the primary direction of
ongoing work.

\noindent\textbf{Benchmark scope.}
The quantitative main-text evidence now spans cantilever scaling, a bridge
hard-problem stress test, and a torsion stress test in which 108--110 of 120
SIMP steps hit the current 1{,}000-iteration CG cap.
The paper therefore still does not establish comparable behavior on the broader
family of standard 3D TO benchmarks, and MBB remains an unresolved cap-limited
case under the current solver configuration.
A companion energy measurement workflow
polls GPU power draw via the standard NVIDIA command-line telemetry utility at
100\,ms intervals and integrates the trace to
yield per-run energy in Joules; the workflow records measured energy alongside the
$\mathrm{TDP} \times t_{\mathrm{wall}}$ upper bound.
These energy rows come from a separate instrumented workflow and should be
compared only within the matched FP64/fused runs of that workflow, not against
the main timing table row-by-row.
The 100\,ms cadence cannot resolve sub-100\,ms power transients, so the
reported Joule values should be read as board-level energy estimates rather
than as cycle-accurate power integrals.
Only the 216\,k and 1\,M cantilever cases were instrumented in that workflow.
Measured board-power traces (SIMP-120 cantilever, RTX\,4090) give
0.648\,Wh for the fused 216\,k run (mean 131.5\,W, 29\% of rated TDP)
versus 2.098\,Wh for the FP64 baseline---a \textbf{3.24$\times$ energy
reduction}.
At 1\,M elements the fused kernel uses 2.979\,Wh (mean 231.7\,W) versus
14.670\,Wh for FP64, a \textbf{4.92$\times$ energy reduction}.
The per-path $\mathrm{TDP} \times t_{\mathrm{wall}}$ upper bounds overestimate
actual measured energy by $1.9$--$3.4\times$ across the four tested
configurations, with the fused 216\,k bound ($2.22$\,Wh vs.\ actual $0.65$\,Wh;
$3.4\times$) being the largest and the fp64 1\,M bound ($28.33$\,Wh vs.\
actual $14.67$\,Wh; $1.9\times$) the smallest, because the GPU operates at
only $29$--$52$\% of its rated 450\,W TDP across the instrumented FP64 and
fused runs.
For the 358\,s fused 2\,M run the TDP upper bound is $\approx$\,45\,Wh, but
extrapolating the measured 231.7\,W mean power gives $\approx$\,23\,Wh.

\noindent\textbf{BF16 arithmetic inapplicable to the CG solve.}
As demonstrated in Section~\ref{sec:results:bf16}, BF16 arithmetic stagnates in
the CG linear solve due to the high condition numbers inherent to SIMP-penalized
stiffness matrices.
The BF16 WMMA kernel is therefore benchmarked for per-matvec throughput only, not
as a usable component of a production SIMP solver.
Integration as a multigrid smoother---where the coarse-grid correction bounds the
effective $\kappa$ seen by the smoother---is identified as the viable path to
extracting BF16 throughput benefit.

\noindent\textbf{Single-GPU scope.}
The present solver operates on the tested single RTX\,4090 consumer GPU.
The largest reported stress test is the 8\,M-element FEA-only solve, which uses
10.56\,GB on the RTX\,4090. Full SIMP runs require additional storage for density
fields, sensitivities, and filter workspace, so practical single-GPU SIMP limits
on the 24\,GB card are lower than a simple FEA-only extrapolation would suggest;
larger problems ultimately require either a larger single GPU or a
distributed-memory multi-GPU approach.
Extending the fused kernel to multi-GPU via NCCL-based halo exchanges and
domain decomposition is left for future work.

\subsection{Path to Geometric Multigrid Integration}
\label{sec:discussion:gmg}

Geometric multigrid (GMG) is the highest-priority solver follow-on to the
present operator paper.
The present fused kernel already provides the fine-grid smoother kernel: each
V-cycle pre/post-smoothing step is a fixed number of matvec-plus-update operations
with the same gather--GEMM--scatter structure.
The additional components required are:
(i) an inter-grid restriction operator ($\mathbf{R}$: inject or full-weighting) that
maps fine-grid residuals to the coarser grid,
(ii) a coarse-grid prolongation operator ($\mathbf{P} = \mathbf{R}^{\top}$),
and (iii) a coarse-grid direct solver (cuSolver or cuSPARSE for the coarsest level).
For the Cartesian structured mesh used in SIMP TO, all three components have closed-form
stencil representations and can be implemented as additional CuPy runtime-compiled
kernel launchers without leaving the Python ecosystem.

Following the classical multigrid treatments of Briggs et al. and Trottenberg
et al.~\cite{briggs2000multigrid,trottenberg2001multigrid} together with the TO-specific
guidance of Peetz and Elbanna~\cite{peetz2021multigrid}, a hybrid GMG--AMG strategy
is planned: GMG for the first 40--60 SIMP iterations (when the topology is diffuse
and the hierarchy is geometrically regular), transitioning to AMG via PyAMG
after the topology has condensed (when AMG's robustness to irregular connectivity
becomes advantageous).
This strategy is intended to substantially reduce the CG workload per SIMP step.
Under the simple assumption that the current 2\,M fused run is dominated by CG work,
that would move the end-to-end wall time materially lower, into a much more favorable
regime than the present Jacobi-PCG path.

Additionally, once GMG bounds the effective condition number at each level below
the BF16 IR threshold, the BF16 WMMA kernel becomes viable
as the V-cycle smoother on the fine grid, potentially recovering the theoretical
tensor-core throughput advantage in a production setting.
This two-path roadmap---GMG for preconditioner improvement and BF16 as a smoother
within the resulting bounded-$\kappa$ subproblem---defines the next phase of
this research program.

\section{Conclusion}
\label{sec:conclusion}

This paper presented four contributions for matrix-free 3D SIMP topology
optimization on the tested RTX\,4090 consumer GPU.

First, we introduced a fused gather--GEMM--scatter CUDA kernel implemented through
CuPy's runtime kernel-compilation interface, preserving a Python-native workflow while eliminating
the intermediate per-element DRAM round-trips of the three-stage baseline. In the reported
study, that fused path achieves 6.0--6.6$\times$ per-matvec speedup (synthetic
hot-path microbenchmark) and 8.9--13.8$\times$ in isolated CUDA-event
measurements on the actual operator (Table~\ref{tab:bw_utilization}), the latter
also reflecting the Python/CuPy dispatch overhead avoided by collapsing three
API calls into one launch;
end-to-end SIMP-120 wall-time speedup is problem-size-dependent: 4.6--7.3$\times$
on cantilever and 4.4$\times$ on the 499{,}125-element torsion benchmark.
Against the same-precision FP32 three-stage path, the fused solver yields
2.3--4.6$\times$ on cantilever and 2.8$\times$ on torsion.
Measured board-power traces show that the fused path also
delivers 3.2--4.9$\times$ energy reduction relative to matched FP64
instrumented runs (0.65--2.98\,Wh vs.\ 2.10--14.67\,Wh at 216\,k and 1\,M
elements), with the GPU operating at 29--52\% of its 450\,W rated TDP across
those instrumented runs.

Second, we reported end-to-end SIMP scaling on the cantilever benchmark together
with a harder torsion stress test and a bridge hard-problem stress test, making
explicit that the paper's SIMP tables use selected valid iterates rather than
last-iterate metrics and that the large FEA-only points are capped stress-test
solves rather than fully converged ones. The bridge family is consistent with the same
FP32 three-stage gain persisting on a distributed-load case, although that
bridge study is not a direct fused-kernel benchmark. MBB remains the unresolved
cap-limited case within the same hard-problem family.

Third, we implemented a BF16 WMMA tensor-core variant of the fused kernel and
showed that the reported 512\,k profiling study includes a
$14.3\times$ BF16-versus-FP64 GEMM proxy timing from a separate PyTorch BF16 GEMM
benchmark, even though the full fused-BF16
matvec remains bounded by the gather/scatter share.

Fourth, we characterized the mixed-precision failure mode of BF16 in the present
Jacobi-preconditioned CG setting. Direct power-iteration estimates of $\kappa(\Kmat)$
reported in Section~\ref{sec:results:kappa} give
$\kappa \approx 6.1\times10^{5}$ at 64\,k, $\approx 1.3\times10^{6}$ at
216\,k, and $\approx 2.3\times10^{6}$ at 512\,k elements, placing
$\varepsilon_{\mathrm{BF16}}\cdot\kappa \approx 2.4\times10^{3}$--$9.1\times10^{3}$ across the three
tested sizes---more than an order of magnitude above the convergence threshold.
These direct measurements show why BF16-IR stagnates: the systems are firmly in
the non-convergence zone, not at the boundary (\Cref{sec:results:kappa}).
That result points toward BF16 as a more plausible multigrid smoother (where
coarse-grid correction bounds the effective $\kappa$ below 256) than as a
drop-in CG inner solve.

Taken together, these results improve on the three-stage Python/CuPy FP64
baseline used as the reference in this study while
also clarifying the present scope. The fused FP32 path is directly supported on
the reported cantilever and torsion studies; the bridge evidence in this paper
is currently FP32/FP64 three-stage only. The current
Jacobi-preconditioned solver still does not establish comparable performance on
harder benchmark families. The implementation is intended to serve as the
foundation for a separate geometric-multigrid solver study needed to close the
remaining iteration-count gap.

\section*{CRediT authorship contribution statement}
Shaoliang Yang: Conceptualization, Methodology, Software, Investigation,
Formal analysis, Visualization, Writing -- original draft. Jun Wang:
Supervision, Conceptualization, Methodology, Writing -- review \& editing,
Funding acquisition. Yunsheng Wang: Validation, Investigation,
Writing -- review \& editing.

\section*{Declaration of competing interest}
The authors declare that they have no known competing financial interests or
personal relationships that could have appeared to influence the work reported
in this paper.

\section*{Data availability}
The full code-and-experiments repository for the present implementation will be
made public at
\url{https://github.com/nbbllxx0/Fused-Gather-GEMM-Scatter-Kernels} once the
arXiv version is online. Until that release, the present preprint records the
methodological details, measurement conventions, and reporting rules needed to
interpret the reported experiments.

\appendix
\section{Reproducibility Notes}
\label{app:repro}
\setlength{\emergencystretch}{2em}%

This appendix records the reader-facing settings needed to interpret the
reported experiments without exposing repository-internal bookkeeping.

\noindent\textbf{Software and hardware context.}
All GPU experiments for the present implementation use the single RTX\,4090
configuration stated in Section~\ref{sec:results:setup}. The local PyTopo3D
comparison in Section~\ref{sec:results:scaling} is a separate CPU baseline
rerun and is not part of that GPU solver path. The reported manuscript values
were generated under Microsoft Windows 10.0.26200.8037 with NVIDIA driver
595.71, Python\,3.11, CuPy\,13.6.0, PyTorch\,2.5.1 built against CUDA\,12.1,
NumPy\,2.2.6, SciPy\,1.15.3, Matplotlib\,3.10.7, scikit-image\,0.25.2,
Pandas\,2.3.3, and PyVista\,0.46.3; the CuPy runtime reports CUDA runtime
version 12090. The GPU used the default 450\,W power limit.
This CUDA\,12.1/12.9 split reflects PyTorch's bundled CUDA\,12.1 user-space
runtime alongside the installed CUDA\,12.9 runtime used by CuPy; that mixed
configuration is supported by the installed NVIDIA driver on this host.
For the contextual PyTopo3D comparison, the deposited CPU rerun log records
20 CPU cores and 69.50\,GB available RAM on the same Windows host, and the
accompanying environment manifest records PyTopo3D 0.1.0, PyPardiso 0.4.7,
and default host scheduling with no explicit thread pinning;
Figure~\ref{fig:external_bars} should therefore be read as contextual timing
evidence rather than as a version-controlled benchmark.
The synthetic hot-path microbenchmark uses the adaptive repeat rule
\texttt{max(100, min(2000, int(5e7 // n\_elem)))} to target about 50\,ms of
total measurement per size.

\noindent\textbf{What is and is not public at the preprint stage.}
The full code-and-experiments repository for the present implementation will be
made public at
\url{https://github.com/nbbllxx0/Fused-Gather-GEMM-Scatter-Kernels} once the
arXiv version is online.
At preprint time, readers should therefore treat this manuscript and
Appendix~\ref{app:provenance} as the authoritative reader-facing record of the
experimental settings, measurement definitions, and reporting conventions used
for the reported results.

\noindent\textbf{Key interpretation rules.}
Table~\ref{tab:bf16conv} reports single cold-start linear solves at uniform
density $\rho=0.5$ and $\penal=3$; they are not full SIMP runs.
Table~\ref{tab:bridge_hard} mixes repeated cold-start FEA means (five timed
calls in the benchmark script's default mode) with single representative
SIMP-60 runs.
The 2\,M, 4.9\,M, and 8\,M FEA-only points are capped-at-1{,}000-iteration
stress-test solves rather than fully converged linear solves.
The FEA scaling-ladder summaries average three timed solves for 216\,k,
512\,k, and 1\,M elements, and two timed solves for 2\,M, 4.9\,M, and 8\,M
elements.
The SIMP scaling rows in Table~\ref{tab:simp_scaling} and
Figure~\ref{fig:simp_scaling} are single representative SIMP-120 runs from the
dedicated scaling workflow, one deposited row per (size, path) pair; repeat
variability is reported separately in Table~\ref{tab:repeat_study}.
The condition-number workflow uses fixed seeds (42 for power iteration and 123
for inverse iteration) and inner SciPy CG solves at $10^{-8}$ relative and
absolute tolerance with a 2{,}000-iteration cap.
The energy numbers come from a separate instrumented workflow and should be
compared only within that workflow's matched FP64/fused runs.
That workflow polls board power through \texttt{nvidia-smi} at an intended
100\,ms interval, starts sampling immediately before the SIMP call, stops
immediately after solver return, and computes energy by trapezoid integration
over the recorded timestamped samples. This cadence cannot resolve sub-100\,ms
power transients, so the resulting Joule values should be read as board-level
energy estimates rather than as cycle-accurate power integrals.

\begin{figure}[!htbp]
  \centering
  \includegraphics[width=0.92\linewidth]{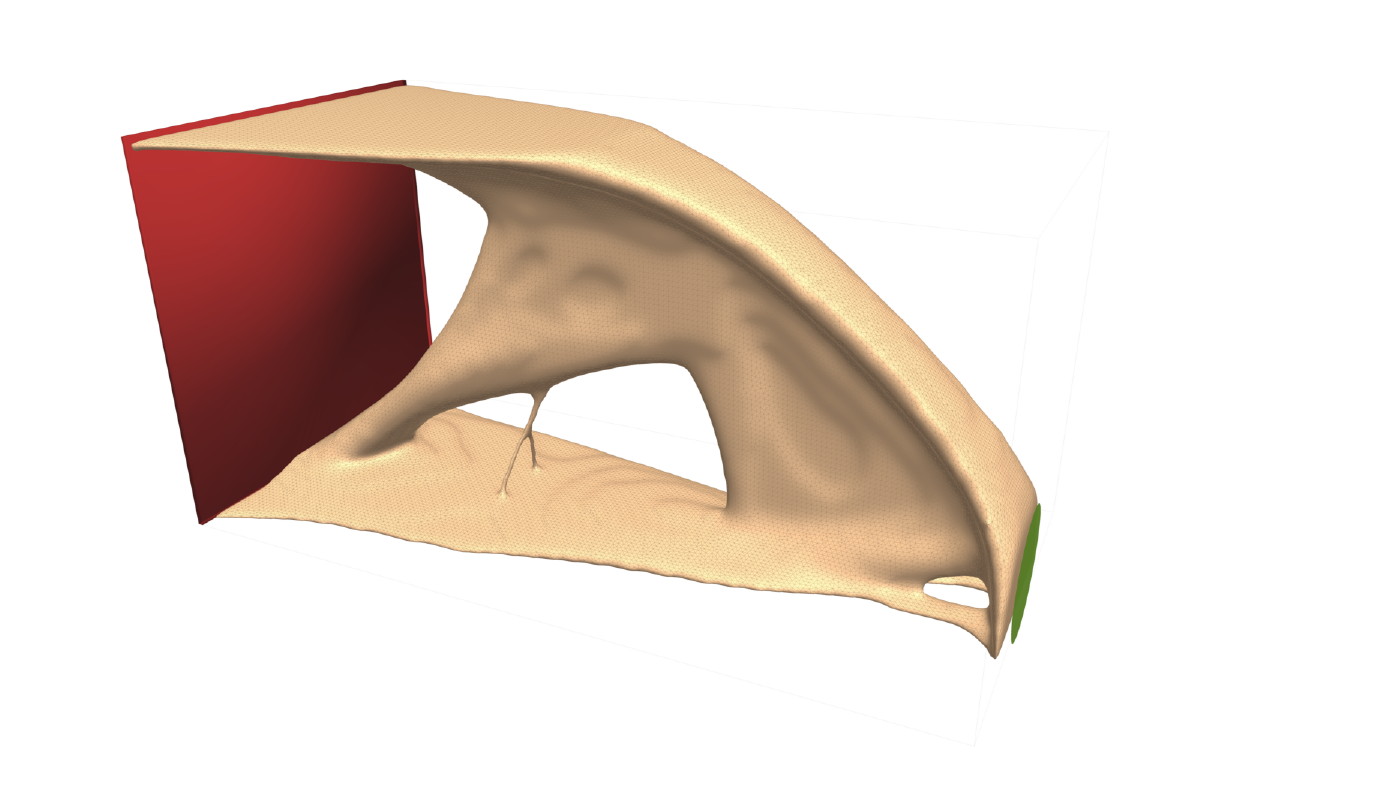}
  \caption{Centered-patch cantilever topology from a separate fused-FP32 SIMP
           rerun outside the quantitative benchmark set. The selected design field
           uses $160\times 80\times 80 = 1{,}024{,}000$ elements on a
           $2\times1\times1$ domain with $\vf = 0.10$, filter radius
           $\rmin = 3.0$, and 100 design iterations. The rendered surface is a
           smoothed marching-cubes isosurface at $\rho=0.5$ from a single fixed
           view (80-step Laplacian smoothing, relaxation factor 0.08); the red
           face indicates the clamped boundary and the green marker
           indicates the centered loaded-face patch on the right face.
           Selected compliance $= 1.830$, selected grayness $= 1.35\times10^{-5}$,
           and the selected design comes from iteration 70 of the 100-step run.}
  \label{fig:centerpatch_cantilever_render}
\end{figure}
\FloatBarrier

\section{Metric Conventions and Artifact Provenance}
\label{app:provenance}
\setlength{\emergencystretch}{2em}%

This appendix states the reporting conventions and the local workflows and
result records used while preparing the main tables and figures in the
preprint.
At preprint time, it serves as the reader-facing provenance record supporting
the main paper.

\noindent\textbf{Selected versus final values.}
The main paper reports \emph{selected} compliance values for the SIMP studies.
Here, ``selected'' means the lowest-compliance iterate among the iterates that
pass the run's validity checks.
In the present implementation, those checks require $\penal \ge 3.0$ and
grayness $g < 0.25$.
These selected values are not guaranteed to be identical to the last iterate of
the reported continuation schedule (120 steps in the main cantilever/torsion
tables and 60 steps in Table~\ref{tab:bridge_hard}).

\noindent\textbf{Figure interpretation.}
Figures~\ref{fig:cantilever_render}, \ref{fig:cantilever_1m_render},
\ref{fig:torsion_render}, and \ref{fig:bridge_render} render selected design
fields as smoothed marching-cubes isosurfaces at $\rho=0.5$, using the render
script's fixed-view layouts and 80-step Laplacian smoothing
($n_{\mathrm{iter}}=80$, relaxation factor $0.08$).
Figure~\ref{fig:bridge_render} is a qualitative companion and is not the source
of Table~\ref{tab:bridge_hard}.
Figure~\ref{fig:external_bars} uses completed 64\,k and 216\,k PyTopo3D rows.

\noindent\textbf{Scope of the validation studies.}
The condition-number tabulation in Section~\ref{sec:results:kappa} covers only
uniform-density states $(\rho=0.5)$ at $\penal\in\{3.0,5.0\}$ for 64\,k,
216\,k, and 512\,k elements.
A separate uniform-density BF16 extension at $\penal=4.5$ supports the brief
remark about higher-penalization stagnation in Section~\ref{sec:results:kappa},
but it is not part of that condition-number table.
The selected-iterate high-cap validation in Section~\ref{sec:results:stats}
covers 216\,k and 512\,k for the FP64 and fused paths.
The 2\,M and 4.9\,M cantilever scaling rows are summary-only representative runs
and are not part of that high-cap validation set.

\noindent\textbf{Repository release status.}
The full code-and-experiments repository for the present implementation will be
made public at
\url{https://github.com/nbbllxx0/Fused-Gather-GEMM-Scatter-Kernels} once the
arXiv version is online.
The list below is therefore the preprint-side provenance record of the local
scripts and result files used during manuscript preparation.

\begingroup
\small
\sloppy
\noindent\textbf{Preprint workflow/result mapping.}
\begin{itemize}[leftmargin=1.3em,itemsep=0.35em,topsep=0.35em]
\item \textbf{Profiling table/figure.}
Generators: the synthetic hot-path profiling workflow and the paper plotting
workflow.
Local result records: the profiling summary CSV and structured JSON export.
Notes: synthetic hot-path microbenchmark matching operator tensor shapes but
using seeded index-pattern probes rather than structured cantilever
connectivity.
\item \textbf{Cantilever FEA scaling figure.}
Generators: the cantilever scaling workflow and the paper plotting workflow.
Local result records: the mid-range, large, and 8\,M FEA scaling summary CSVs.
Notes: cold-start FEA rows at uniform density $\rho=0.5$; the 2\,M, 4.9\,M,
and 8\,M rows are cap-limited stress-test solves.
\item \textbf{Cantilever SIMP scaling table/figures.}
Generators: the cantilever scaling workflow and the paper plotting workflow.
Local result records: the mid-range, 1\,M, and 2\,M/4.9\,M SIMP scaling
summary CSVs.
Notes: representative single-run SIMP-120 scaling rows; reported compliance is
the selected best-valid compliance, not the last iterate.
\item \textbf{External CPU comparison figure.}
Generators: the external CPU rerun comparison workflow and the paper plotting
workflow.
Local result records: the PyTopo3D-versus-GPU comparison CSV, the deposited
PyTopo3D CPU log, the dedicated 64\,k fused rerun CSV, and the mid-range SIMP
scaling CSV.
Notes: contextual timing comparison only; the 64\,k fused bar comes from a
separate deposited 64\,k rerun, and the PyTopo3D record is paired with the
environment manifest that records PyTopo3D 0.1.0, PyPardiso 0.4.7, and
default host scheduling with no explicit thread pinning.
\item \textbf{Torsion table/trajectory figure.}
Generators: the torsion benchmark workflow and the paper plotting workflow.
Local result records: the torsion summary CSV and the FP64, FP32, and
fused iteration-history JSON files.
Notes: 108--110 of the 120 SIMP steps hit the current 1{,}000-iteration CG
cap across the three deposited histories.
\item \textbf{MBB/bridge hard-problem table.}
Generator: the hard-problem benchmark workflow.
Local result record: the hard-problem summary CSV.
Notes: the MBB and bridge rows cover FP64 and FP32 three-stage paths only;
fused is not benchmarked in this table.
\item \textbf{BF16 convergence table.}
Generator: the BF16 iterative-refinement smoke-test workflow.
Local result records: the BF16 smoke-test CSV and JSON outputs.
Notes: single cold-start linear solves at uniform density $\rho=0.5$ and
$\penal=3$; not full SIMP runs.
\item \textbf{Condition-number study.}
Generator: the condition-number estimation workflow.
Local result records: the condition-number CSV and JSON outputs.
Notes: uniform-density states only ($\rho=0.5$; $\penal\in\{3,5\}$); late-SIMP
conditioning statements remain explicitly inferential.
\item \textbf{Repeat, determinism, high-cap, and energy studies.}
Generators: the repeat-study, determinism, high-cap validation, and energy
benchmark workflows.
Local result records: the repeat-study CSV/JSON pair, determinism CSV/JSON
pair, fully converged CSV/JSON pair, and energy CSV/JSON pair.
Notes: supporting studies for variability, atomic-scatter determinism,
selected-iterate validity, and power-trace energy.
\item \textbf{Residual figure.}
Generators: the residual-capture workflow and the paper plotting workflow.
Local result record: the CG residual-history JSON file.
Notes: representative cold-start cantilever solve at uniform density
$\rho=0.5$.
\item \textbf{Qualitative render panels.}
Generator: the topology-render workflow.
Local result records: the selected-density arrays and metadata sidecars for
the 216\,k cantilever, 1\,M cantilever, 499{,}125-element torsion, 216\,k bridge, and
1\,M low-volume-fraction corner-load examples.
Notes: all qualitative panels use selected-design density fields rendered as
smoothed marching-cubes isosurfaces at $\rho=0.5$.
The metadata sidecars record the selected-run/source metadata, while the
fixed camera layouts and panel compositions are defined in the render script.
\end{itemize}
\endgroup

\FloatBarrier

\bibliographystyle{unsrtnat}
\bibliography{refs/references}

\end{document}